\documentclass[usenatbib]{mn2e}
\setcitestyle{notesep={; }}
\usepackage{graphicx}
\usepackage{amsmath}
\usepackage{epsfig}
\usepackage[bookmarks=true,bookmarksnumbered=true,bookmarksopen=false]{hyperref}
\usepackage{comment}

\newcommand\msun{{\, \rm M_\odot}}
\newcommand\zsun{{\, \rm Z_\odot}}
\newcommand\pc{{\, \rm pc}}

\newcommand\yr{{\, \rm yr}}
\newcommand\myr{{\, \rm Myr}}

\def\gtsima{$\; \buildrel > \over \sim \;$}
\def\ltsima{$\; \buildrel < \over \sim \;$}
\def\gtrsim{\lower.5ex\hbox{\gtsima}}
\def\lesssim{\lower.5ex\hbox{\ltsima}}

\title[Mass loss versus dynamical heating in star clusters]
{The impact of metallicity-dependent mass loss versus dynamical heating on the early evolution of star clusters}

\author[Trani, Mapelli \&{} Bressan]
{A. A. Trani$^{1,2}$, M. Mapelli$^{2}$, \&{} A. Bressan$^{1,2}$
\\
$^1$Scuola Internazionale Superiore di Studi Avanzati (SISSA), Via Bonomea 265, I--34136, Trieste, Italy; {\tt aatrani@gmail.com}\\
$^2$INAF-Osservatorio Astronomico di Padova, Vicolo dell'Osservatorio 5, I--35122, Padova, Italy
}

\begin{document}

\date{}

\maketitle

\begin{abstract}

We have run direct N-body simulations to investigate the impact of stellar evolution and dynamics on the structural properties of young massive ($\sim 3\times 10^4 \msun$) star clusters (SCs) with different metallicities ($Z=1, 0.1, 0.01 \zsun$). Metallicity drives the mass loss by stellar winds and supernovae (SNe), with SCs losing more mass at high metallicity. We have simulated three sets of initial conditions, with different initial relaxation timescale.
We find that the evolution of the half-mass radius of SCs depends on how fast two-body relaxation is with respect to the lifetime of massive stars. If core collapse is slow in comparison with stellar evolution, then mass loss by stellar winds and SNe is the dominant mechanism driving SC evolution, and metal-rich SCs expand more than metal-poor ones. In contrast, if core collapse occurs on a comparable timescale with respect to the lifetime of massive stars, then SC evolution  depends on the interplay between mass loss and three-body encounters: dynamical heating by three-body encounters (mass loss by stellar winds and SNe) is the dominant process driving the expansion of the core in metal-poor (metal-rich) SCs. As a consequence, the half-mass radius of metal-poor SCs expands more than that of metal-rich ones. 
 We also find core radius oscillations, which grow in number and amplitude as metallicity decreases. 
\end{abstract}


\section{Introduction}

Dense star clusters (SCs) are collisional systems: their two-body relaxation timescale is shorter than their lifetime. This causes the evaporation of stars from the core, removing kinetic energy. Since a self-gravitating system has a negative heat capacity, the velocity dispersion of the core increases as it contracts. More stars escape from the core, which loses even more kinetic energy. This runaway process is called gravothermal instability and leads the core to collapse  \citep[e.g.][]{spitzer87,gravmill,bin&tre}.


Only an energy source in the core can halt the collapse and quench the instability. This energy source can be represented by three-body encounters, i.e. close encounters between a binary and a single star. 
During such encounters, part of the internal energy of the binary may be redistributed as kinetic energy between the single star and the centre of mass of the binary. In this way, the binary hardens (i.e. its binding energy increases) and the kinetic energy of the system increases \citep{heggie75}. 
By this process, called binary hardening, few binaries in the core can provide the kinetic energy needed to restore the virial equilibrium and reverse the core collapse. If there are no primordial binaries in the core, binary formation is triggered by the high stellar density of the core during the collapse. In the post-collapse phase, the energy generated by three-body encounters in the core is driven outwards by two-body relaxation and the SC expands. This expansion causes the half-mass radius to increase according to $r_{\rm hm} \propto t^{2/3}$ \citep{elson87}.

Mass loss by stellar evolution can deeply affect the evolution of a SC before and after core collapse \citep{a&g77,a&g80,applegate86,chernoff&90,port07,vesp&09,lamers10,gieles13}. 
Moreover, supernovae (SNe) occur in the first 50 Myr since the birth of a SC. 

Metallicity ($Z$) also plays a relevant role, since it determines the efficiency of stellar winds \citep{leitherer&92,maeder92,pols&98,portinari&98,kudritzki02}. 
High-metallicity stars lose more mass by stellar winds than low-metallicity stars \citep{vink01,vink05}. Since it drives the mass loss rate of a star, metallicity indirectly affects the outcome of a SN explosion.

\citet{schul12} performed N-body simulations of intermediate-mass young SCs with a wide spectrum of metallicities, 
 and found the size of SCs to be metallicity dependent. Metal-rich SCs expand more rapidly than metal-poor SCs in the first 20 Myr, while the trend reverses thereafter. 
Similarly, \citet{down12} simulated globular clusters with different metallicity using Monte Carlo methods. He found that the half-mass radius of metal-poor SCs is smaller at early times than that of metal-rich SCs, but then grows larger within a relaxation timescale, in agreement with \citet{schul12}.

\citet{m&b13} ran  N-body simulations of intermediate-mass young SCs with different metallicity, and found that the half-mass radius of metal-poor SCs grows larger than that of metal-rich SCs, while the core radius of  metal-poor SCs expands less than that of metal-rich SCs after core collapse. They interpreted this result as an effect of the interplay between mass loss by stellar winds and dynamical heating, the expansion of the core being driven mostly by mass loss in metal-rich SCs and by three-body encounters in metal-poor SCs.

On the other hand, \citet{sippel12} investigated the effect of metallicity on massive $N=10^5$ SCs using direct N-body simulations with stellar and binary evolution (\citet{hurley00,hurley02}). They found no structural differences between SCs at different metallicities. 

Our aim is to check the relative importance of binary hardening and metallicity-dependent stellar evolution in determining the structural properties of SCs. In particular, we will expand and generalize the results presented in \citet{m&b13}, by considering a different SC mass range, different central densities and concentrations.

In Section~\ref{sec:method}, we describe the methodology that we employed for our simulations; in Section~\ref{sec:results} we present our results, with particular attention for the evolution of core and half-mass radius, and for the core radius oscillations. In Section~\ref{sec:discussion}, we discuss the implications of our work and we compare it with analytic models. Our conclusions are presented in  Section~\ref{sec:conclusions}.

\begin{figure}
  \begin{center}
    \epsfig{figure=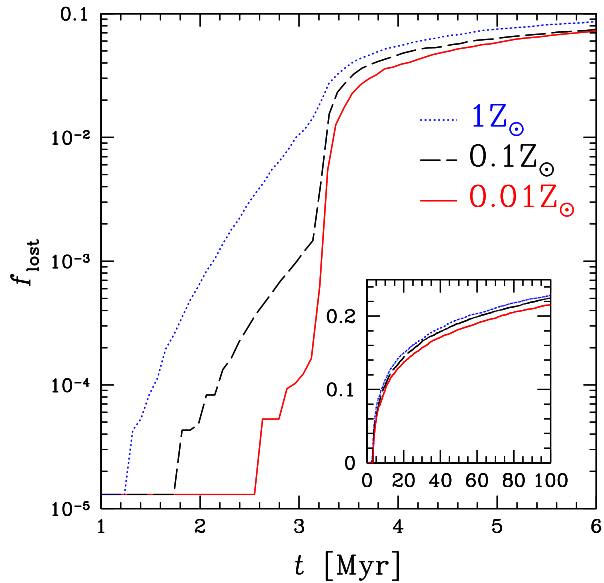,width=8.0cm} 
  \end{center}
    \caption{Cumulative mass loss by stellar winds and SNe normalised to the initial mass of the SC as a function of time for three different metallicities. Solid red line: $Z=0.01 \zsun$; dashed black line: $Z=0.1 \zsun$; dotted blue line: $Z=1 \zsun$. Each line is the median value of 10 simulated SCs for different metallicity.
}
\label{fig:1}
\end{figure}

\section{Method}\label{sec:method}
The simulations were run with the {\sc starlab} software environment \citep[see also \citealt{port96}]{port01}, which uses a fourth-order Hermite integrator to compute the dynamics of stars and binaries. Single star and binary evolution are implemented in the {\sc SeBa} routine \citep{port96}. Our version of {\sc starlab} includes new recipes for stellar evolution, as described in \citet{map&13}. In particular, it includes the metallicity-dependence of stellar radius, temperature and luminosity, by implementing the polynomial fitting formulae of \citet{hurley00}. It also includes updated recipes for mass loss by stellar winds of main sequence stars, by using the prescription of \citet{vink01}.

\citet{map&13} also added an approximate treatment for luminous blue variable (LBV) and Wolf-Rayet (WR) stars. In this version of {\sc starlab}, helium giants coming from stars with $m_{\rm zams} > 25 \msun$ are labelled as WR stars and they undergo mass loss-rate by stellar winds given by the formula: $\dot m = 10^{-13} \left(L/{\rm L_\odot}\right)^{1.5}\left(Z/{\rm Z_\odot}\right)^{0.86} \msun \yr^{-1}$, where $\rm L_\odot$ is the solar luminosity \citep{map&13,b10,vink05,ham98}.

In the code, post-main-sequence stars with luminosity $L > 6\times 10^5 \, \rm L_\odot$ and radius $R > 10^5 \left(L/{\rm L_\odot}\right)^{-0.5} R_\odot$, where $R_\odot$ is the solar radius, are labelled as LBV stars \citep{hump94}. Their mass-loss rate is calculated as $\dot m = f_{\rm LBV} \times 10^{-4} \yr^{-1}$, where $f_{\rm LBV}=1.5$ is an arbitrary constant chosen to reproduce the most massive known stellar BHs \citep{b10}.

Stellar winds of asymptotic giant branch stars are modelled as in the original version of {\sc starlab} and do not include any metal-dependent recipes. We assume that the mass lost by stellar winds and SNe is ejected from the SC, and it is thus removed from the simulation. This assumption is realistic for SN ejecta and also for the winds of massive stars, which are expected to move fast ( $> 2000$ km s$^{-1}$ for the O stars, e.g. \citealt{muijres12}; $>1000$ km s$^{-1}$ for the WR stars, e.g. \citealt{vink05}; \citealt{martins08}) with respect to the central escape velocity of the simulated SCs ($\sim{}12$ km s$^{-1}$ for SCs of set A and B, and $\sim{}6$ km s$^{-1}$ for SCs of set B). Stellar winds by AGB stars have much smaller velocities ($10-20$ km s$^{-1}$, \citealt{loup93}; \citealt{gonzalez03}; \citealt{nanni13}; \citealt{schoier13}), but still sufficiently high to escape from our simulated SCs. Furthermore, AGB stars do not play an important role for the results presented in this paper, as the winds from AGB stars become important at $\gtrsim{}50$ Myr.

The formation of stellar remnants is implemented as described in
\cite{map&13}. In particular, black hole (BH) masses for various metallicities follow the
distribution described in Fig. \ref{fig:1} of Mapelli et al. (2013, see also \citealt{fryer99}; \citealt{fryer01}; \citealt{b10}; \citealt{fryer12}). If the final mass $m_{\rm fin}$ of
the progenitor star (i.e. the mass before the collapse) is $> 40$ M$_\odot$ , we assume that the SN fails and that
the star collapses quietly to a BH. The requirement that  $m_{\rm fin}> 40$ M$_\odot$ implies that only stars with ZAMS mass $\ge{}80$ and $\ge{}100$ M$_\odot$ 
can undergo a failed SN at $Z = 0.01$ and 0.1 Z$_\odot$ , respectively. 
If  $m_{\rm fin}> 40$ M$_\odot$, the mass of the BH is derived as $m_{\rm BH} = m_{\rm CO} + f_{\rm coll}\,{} ( m_{\rm He} + m_{\rm H} )$, where $m_{\rm CO}$ is the final mass
of the Carbon Oxygen (CO) content of the progenitor, while $m_{\rm He}$ and $m_{\rm H}$ are the residual mass of Helium (He) and of Hydrogen (H),
respectively.  $f_{\rm coll}$ is the fraction of He and H mass that collapses
to the BH in the failed SN scenario. We assume $f_{\rm coll} = 2 / 3$ to match the maximum values of $m_{\rm BH}$ at $Z=0.01$ Z$_{\odot}$ derived by \cite{b10}. In this scenario, BHs with mass up to $\sim{} 80$ M$_\odot$ ( $\sim{}40$ M$_\odot$) can form if the metallicity of the progenitor is $Z \sim{}0.01$ Z$_\odot$ ( $Z\sim{}0.1$ Z$_\odot$). BHs that form from quiet collapse are assumed to receive no natal kick (\citealt{fryer12}). For BHs that form from a SN explosion, the natal kicks were drawn from the same distribution as neutron stars but scaled with the ratio of the mass (see \citealt{map&13} for details).


\subsection{Initial conditions and simulation grid}

The SCs are initialised as a multi-mass, isotropic \citet{king66} model composed of $N=50000$ stars and no initial binaries. Neglecting the primordial binaries  increases the importance of formation and hardening of binaries by three-body encounters during core collapse, without altering the behaviour of core and half-mass radius significantly, as shown by \citet{m&b13}. It also diminishes the statistical noise of the simulations, allowing to determine more easily the moment of core collapse \citep{heggie06}. We adopt two values for the dimensionless potentials: $W_{0}=5$ and $W_{0}=9$, which correspond to an initial concentration of $c = 1.031$ and $c = 2.120$, respectively. Two different virial radii $r_{\rm vir}$ were chosen for the models with $W_{0}=5$: $r_{\rm vir}=1 \pc$ (set A) and $r_{\rm vir}=5 \pc$ (set B). The SCs with $W_{0}=9$ were modelled with $r_{\rm vir}=1 \pc$ only (set C). The stars follow a \citet{kroupa01} IMF with $m_{\rm max}=150 \msun$ and $m_{\rm min}=0.1 \msun$. The values of the initial relaxation timescale are listed in Table~\ref{tab:ic}, along with the main initial conditions. These initial conditions resemble the properties of observed young massive SCs \citep[see e.g.][]{ymsc10}. 

Each model was run for three metallicities: $Z=1, 0.1, 0.01 \zsun$. For these initial conditions, SCs with different metallicity experience different mass loss by stellar evolution, as shown in Fig. \ref{fig:1}. Regardless of the metallicity, most of the mass loss by stellar evolution occurs in the first $\simeq 10 \myr$, so we expect that stellar mass loss will drive the dynamics only in the early evolution of the SCs. In the following, we will define $t_{\rm se}\simeq 6 \myr$ the lifetime of the massive stars ($>30$ M$_\odot$).

We have run each simulation for at least 100 Myr. We ran 10 realisations for each set of initial conditions, changing only the random numbers used to compute each realisation. 
We checked that there are no significant differences in the median values of core and half-mass radius if we consider either five or ten realizations. Thus, ten realizations per metallicity are sufficient to filter out most stochastic fluctuations. No external tidal field was set for the simulations. In this way we focus on the intrinsic properties of the simulated SCs.

\begin{table}
\begin{center}
\caption{Initial conditions.}\label{tab:ic} \leavevmode
\begin{tabular}[!h]{c c c c c c c c c}
\hline
Set
& $N$ 
& $M$ 
& $r_{\rm vir}$ 
& $W_0$ 
& $\rho_{\rm c}$ 
& $t_{\rm rh}$ 
& $t_{\rm rc}$
\\
&
& (10$^4$ M$_\odot$)
& (pc)
& 
&  (M$_\odot$ pc$^{-3}$)
& (Myr)
& (Myr)
\\
\hline
\noalign{\vspace{0.1cm}}
A & 50K & 3.25 & 1 & 5 & $2\times 10^4$ & 36 & 29 \\
B & 50K & 3.25 & 5 & 5 & $2\times 10^2$ & 394 & 308 \\
C & 50K & 3.25 & 1 & 9 & 10$^7$ & 44 &  $\lesssim 1$ \\

\hline
\end{tabular}
\begin{flushleft}
\footnotesize{
$N$: number of centres of mass; 
$M$: SC average mass; 
$W_0$: dimensionless central potential of the \citet{king66} model; 
$\rho_{\rm c}$: core mass density; 
$t_{\rm rh}$: relaxation timescale at half-mass radius\footnotemark[1] in $\myr$;
$t_{\rm rc}$: relaxation timescale at core radius\footnotemark[2] in $\myr$;}
\end{flushleft}
\end{center}
\end{table}

\begin{figure*}
  \begin{center}
    \epsfig{figure=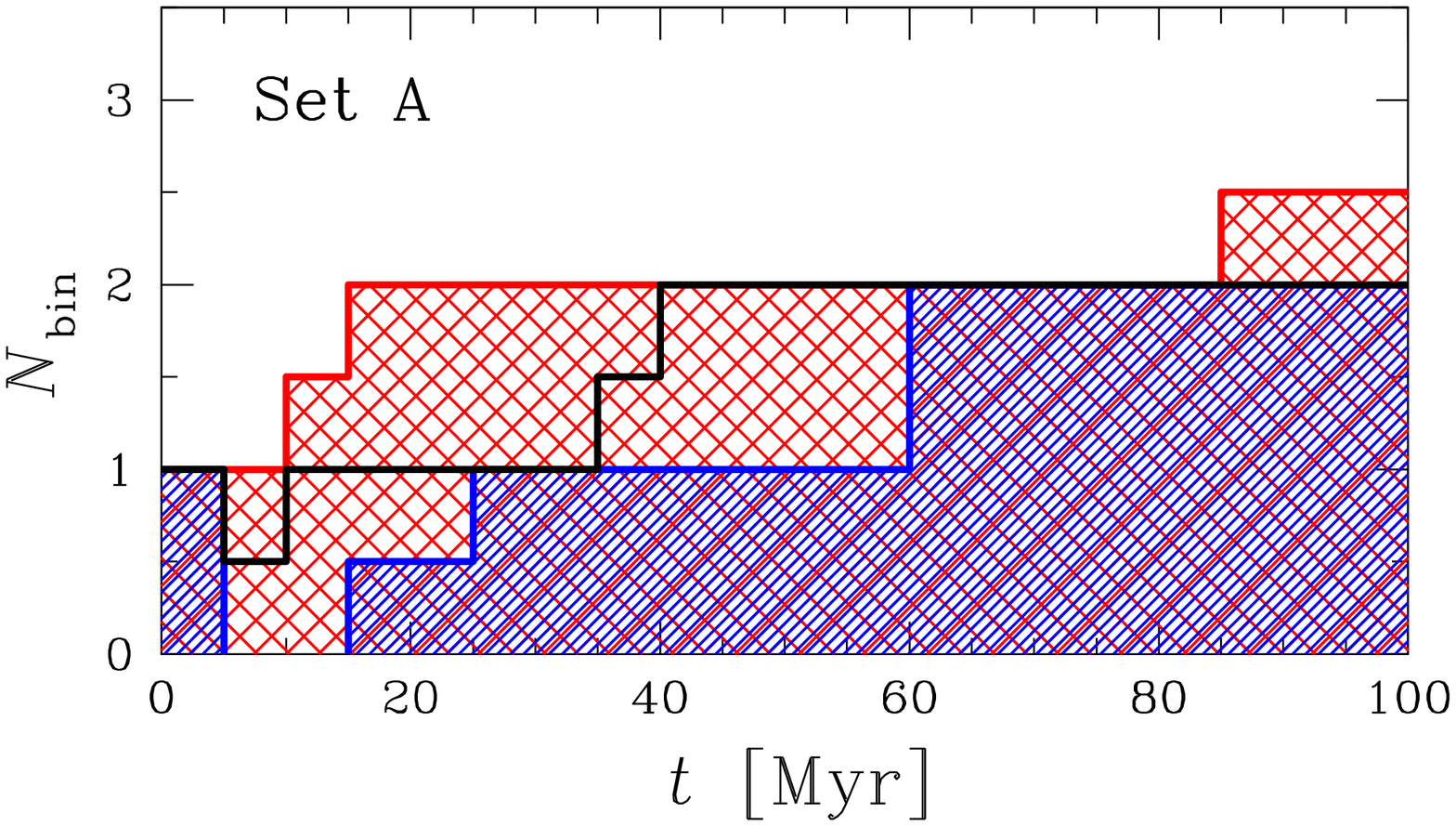,width=5.8cm} 
    \epsfig{figure=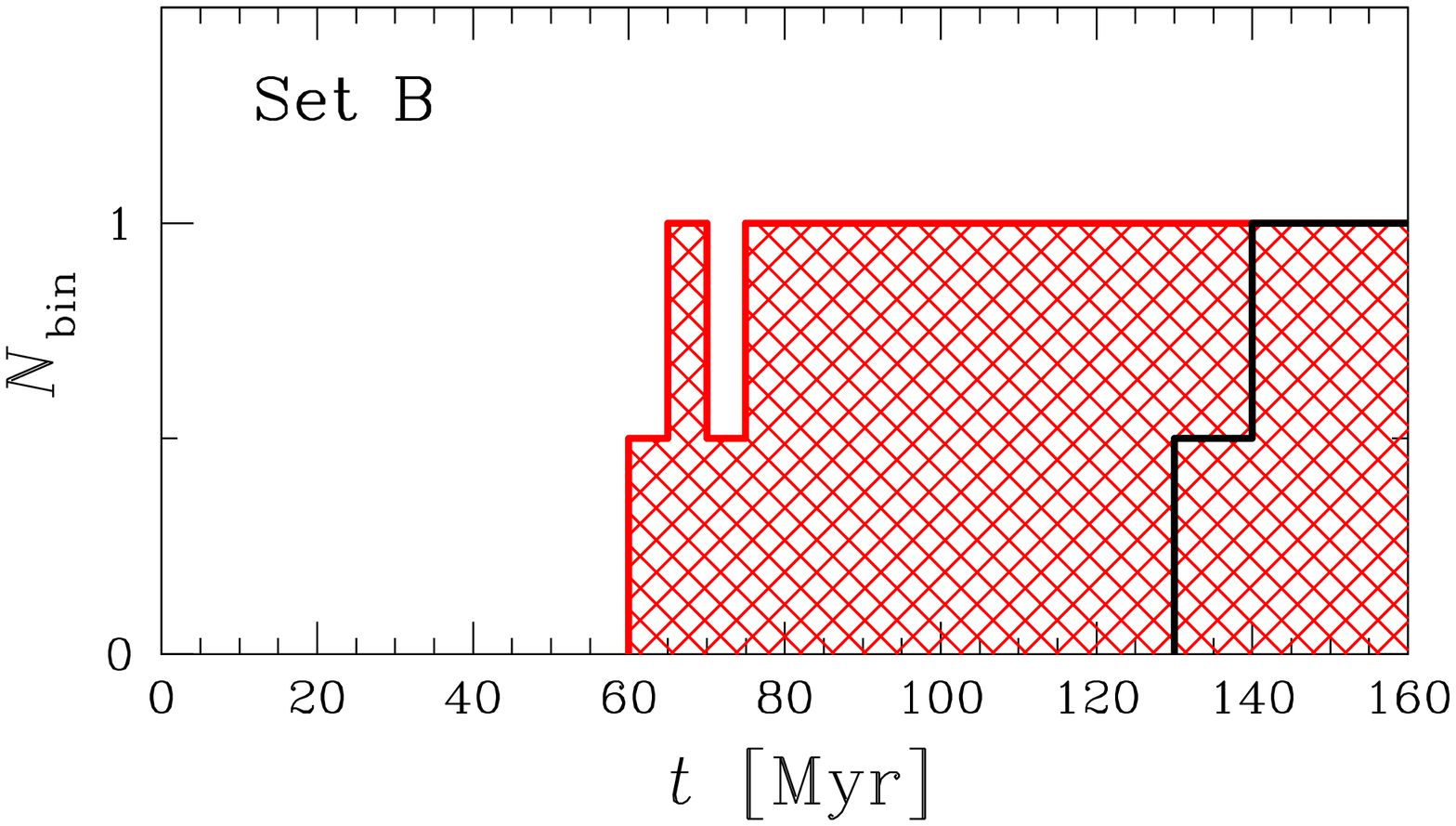,width=5.8cm} 
    \epsfig{figure=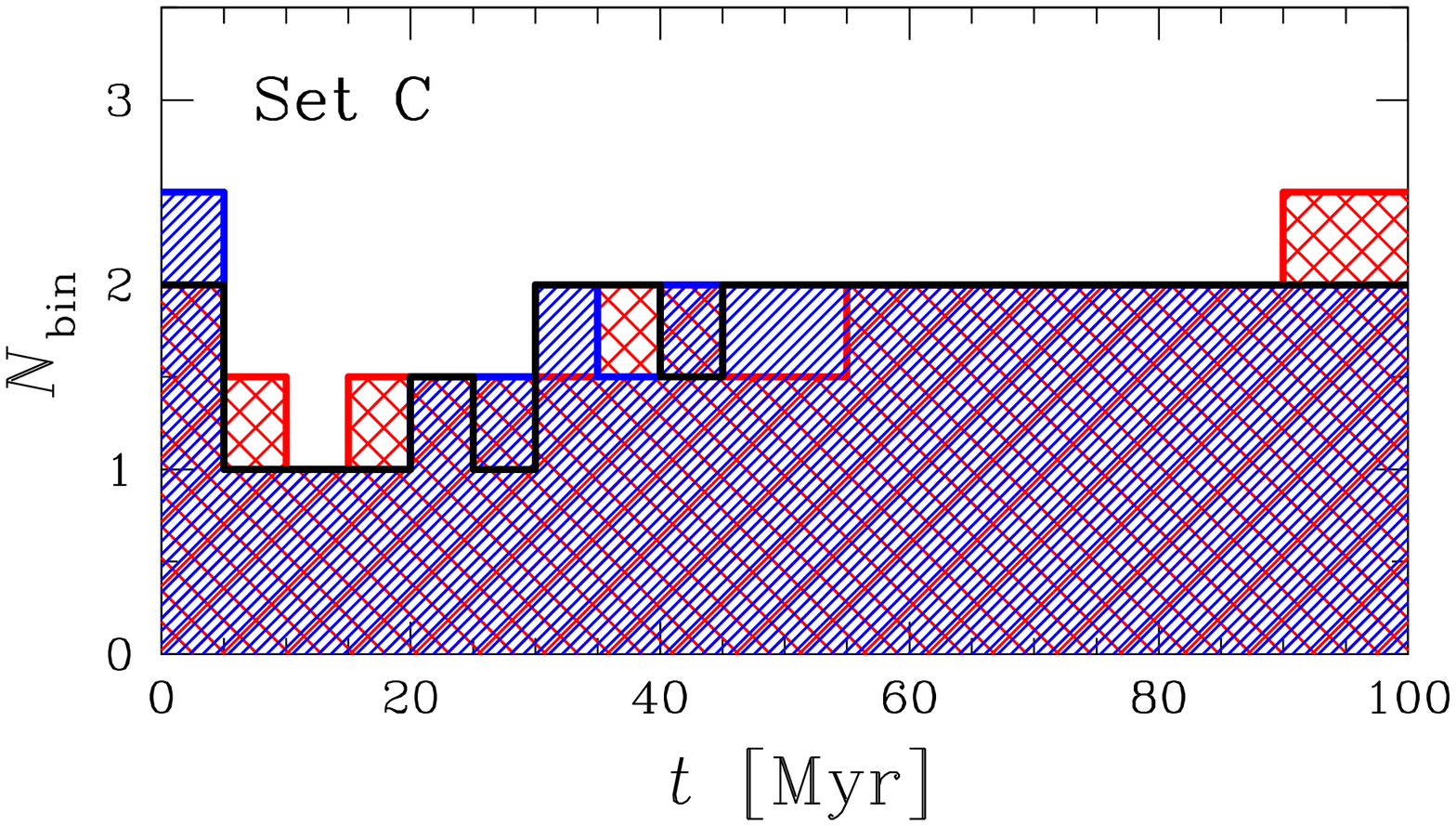,width=5.8cm} 
  \end{center}
  \caption{Number of hard binaries in our simulations as a function of time for the three considered metallicities: $Z=1\zsun$ (hatched blue histogram), $Z=0.1\zsun$ (black empty histogram) and $Z=0.01\zsun$ (cross-hatched red histogram). Since our simulations do not have primordial binaries, all the binaries are formed dynamically. Left-hand panel: SCs with initial $r_{\rm vir}= 1 \pc$ and $W_0=5$ (set A). Middle panel: SCs with initial $r_{\rm vir}= 5 \pc$ and $W_0=5$ (set B). Right-hand panel: SCs with initial $r_{\rm vir}= 1 \pc$ and $W_0=9$ (set C). Each histogram is the median value of 10 realisations.}
\label{fig:2}
\end{figure*}

\footnotetext[1]{Computed with the formula:\\ $t_{\rm rh} \simeq 0.19 \myr \,(r_{\rm hm}/1 \pc)^{3/2} (M/1 \msun)^{1/2} (\langle m \rangle/1 \msun)^{-1}$\\ where $r_{\rm hm}$ is the half-mass radius, $M$ is the total mass of the SC and $\langle m \rangle$ is the mean stellar mass.}
\footnotetext[2]{Computed with the formula $t_{\rm rc} = \frac{\sigma_{\rm c}^{3/2}}{15.4 \,\mathrm{G}^2 \langle m \rangle \rho_{\rm c} \ln\Lambda}$, where $\sigma_{\rm c}$ is the three-dimensional velocity dispersion of the core, $\rho_{\rm c}$ is the core density, $\mathrm{G}$ is the gravitational constant and $\ln\Lambda$ is the Coulomb logarithm. In this paper we set $\ln\Lambda=10$.}

\section{Results}\label{sec:results}

\subsection{Evolution of core and half-mass radius}

In this section we discuss the structural evolution of the simulated SCs. All the quantities discussed in this section are median quantities computed from the ten realisation for each metallicity.
To check if binary hardening is driving the reverse of core collapse we follow the binding energy of the binaries in the simulations. Since the simulations have no primordial binaries, the total binary binding energy at a given time corresponds to the kinetic energy injected into the SC.
We also checked the number of binaries formed during the simulations. Fig. \ref{fig:2} shows the number of binaries as a function of time for set A, set B and set C. No more than three hard binaries are present on average at a single time, indicating that most of the binary binding energy is retained in few hard binaries. The number of binaries formed during the simulations depends on the considered initial conditions (set A, B and C), and on the metallicity.

\subsection*{Set A  }

SCs of set A collapse at the same time, regardless of the metallicity. At $t_{\rm cc} \simeq 3$ Myr, the collapse is halted, the core bounces and begins to expand. Metallicity affects only the post-collapse phase: the core bounce is stronger at higher metallicity (Fig. \ref{fig:3}, left-hand panel). This difference is maximum at $10 \myr$, when the mass loss from the most massive stars is over, and the core radius of $Z = 1 \zsun$ SCs has grown 60 per cent larger than that of $Z=0.01 \zsun$ SCs. We also find that, in the long run, the core radius of metal-poor SCs becomes larger, on average, than the core radius of $Z=1 \zsun$ SCs.

The half-mass radius of metal-poor SCs expands more than that of metal-rich ones (Fig. \ref{fig:3}, middle panel). At 100 Myr, the half-mass radius of $Z=0.01 \zsun$ SCs is 14 per cent larger than the half-mass radius of $Z=1 \zsun$ SCs. The reasons for the expansion of the halo will be discussed in Section~\ref{sec:discussion}.

The right-hand panel of Fig. \ref{fig:3} shows the evolution of the binary binding energy. The first peak in the binary binding energy coincides with the core bounce, but disappears immediately after for SCs with $Z=1 \zsun$ and $Z=0.1 \zsun$. Only in the case with $Z=0.01 \zsun$ the hardening goes on right after the bounce at 3 Myr. In general, the hardening of binaries starts later at higher metallicity. In SCs with $Z=1 \zsun$, binary hardening begins at 20--50 Myr, depending on the simulation. In SCs with $Z=0.1 \zsun$, the binary hardening occurs even earlier, at 10--25 Myr.
The left-hand panel of Fig. \ref{fig:2} shows that only two hard binaries form on average during each simulation of set A. Binaries are formed earlier in metal-poor SCs, as a consequence of the earlier binary hardening.


\subsection*{Set B }
The left-hand panel of Fig. \ref{fig:4} shows the evolution of the core radius of the SCs of set B. During the first 3 Myr there is a weak decrease in the core radius. 
 The decrease of core radius is the beginning of a long and slow phase of core collapse, interrupted by the first SN explosions at 3 Myr. The impulsive mass loss at 3 Myr causes an expansion of both core and half-mass radius. The initial expansion of the core is over at $\approx 7 \myr$. Then, the core begins to collapse, faster at lower metallicity. In the simulations with $Z=1 \zsun$ the core radius remains approximately constant, and only three SCs  out of ten show a decrease in core radius after 120--130 Myr. Six of the SCs with $Z=0.1 \zsun$ show a core bounce at $\approx$160 Myr, while three experience an early core collapse at 100--120 Myr. In SCs with $Z=0.01 \zsun$ the collapse begins immediately after the initial expansion, but the time when core collapse stops varies from SC to SC, and goes from 50 Myr to 140 Myr.

At 3 Myr the half-mass radius evolves in the opposite way with respect to SCs of set A: the half-mass radius of metal-rich SCs expands more than that of the metal-poor SCs, at least during the first 60 Myr (Fig. \ref{fig:4}, middle panel).
The half-mass radius of metal-rich SCs at $\approx$35 Myr is 5 per cent larger than that of metal-poor SCs. 

At 60--70$\myr$ the half-mass radius of SCs with $Z=0.01 \zsun$ begins to expand faster than that of metal-rich SCs. This coincides with the beginning of binary hardening in SCs with $Z=0.01 \zsun$. The half-mass radius of SCs with $Z=0.01 \zsun$ at 160 Myr has grown 4 per cent larger than the half-mass radius of metal-rich SCs. 
With respect to the SCs of set A, the difference in half-mass radius among SCs of different metallicity remains $<10$ per cent throughout the simulations.

Binary hardening occurs earlier at low metallicity (Fig. \ref{fig:4}, right-hand panel). Binary hardening is absent in the SCs with $Z=1 \zsun$, since they are the only SCs not experiencing core collapse in the time spanned by the simulations. In fact, no hard binaries are formed in $Z=1 \zsun$ SCs, and only one is formed (on average) in SCs with $Z\le{}0.1 \zsun$ (Fig. \ref{fig:2}, middle panel).

\subsection*{Set C }
The high ($W_0=9$) initial concentration of SCs of set~C implies that the core of the SCs of set C is already collapsed at the beginning of the simulations. During the simulations, the core radius never reaches values as small as the initial one. Since the core is already collapsed, strong three-body encounters immediately occur and cause the core to rapidly expand (Fig. \ref{fig:5}, left-hand panel). The right-hand panel of Fig. \ref{fig:5} shows a peak of binary binding energy at 3 Myr, which is five orders of magnitude higher than the one in the less concentrated SCs of set A.

At $\approx$5 Myr the core radius of SCs with $Z = 1, 0.1 \zsun$ has expanded 30 per cent more than that of $Z = 0.01 \zsun$ SCs. This difference is half as much as in the SCs of set A.
Afterwards, the expansion slows down due to the diminished stellar mass loss rate. However, the core density in metal-poor SCs is still high enough to make binary hardening go on. As in SCs of set A, we find that the core of metal-poor SCs expands faster than the core of metal-rich SCs, in the long run.

\begin{figure*}
  \begin{center}
    \epsfig{figure=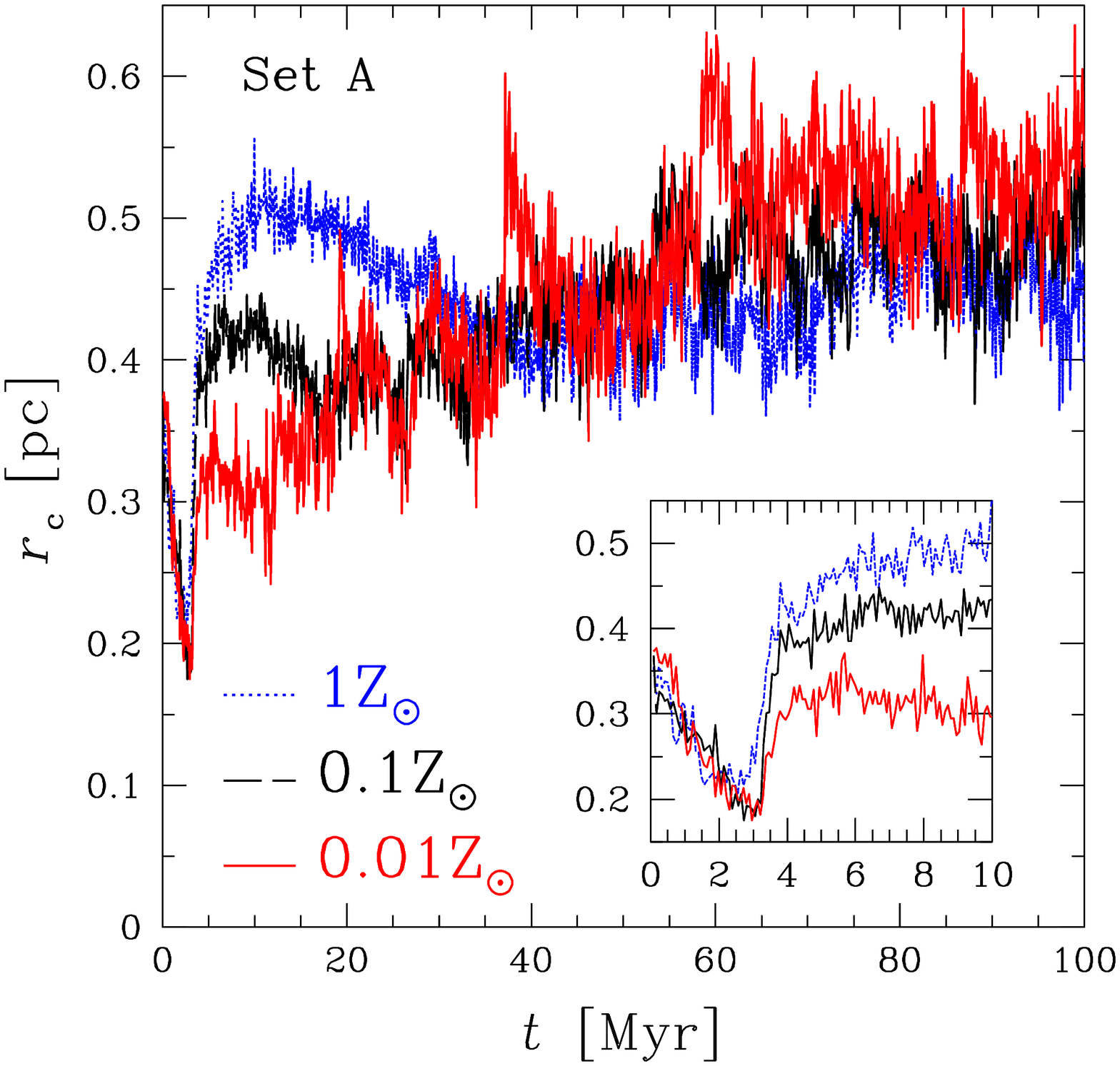,width=5.8cm} 
    \epsfig{figure=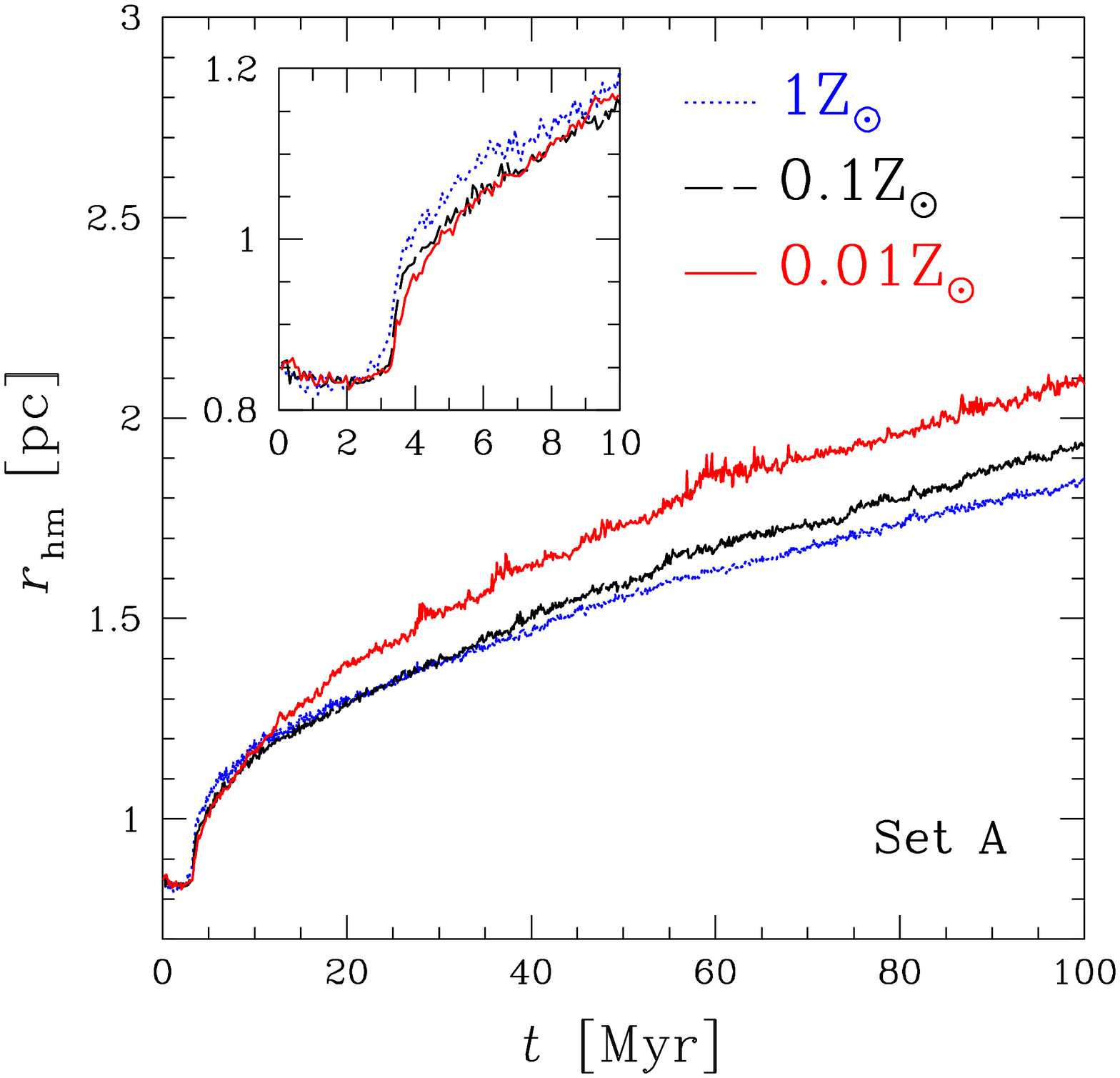,width=5.8cm} 
    \epsfig{figure=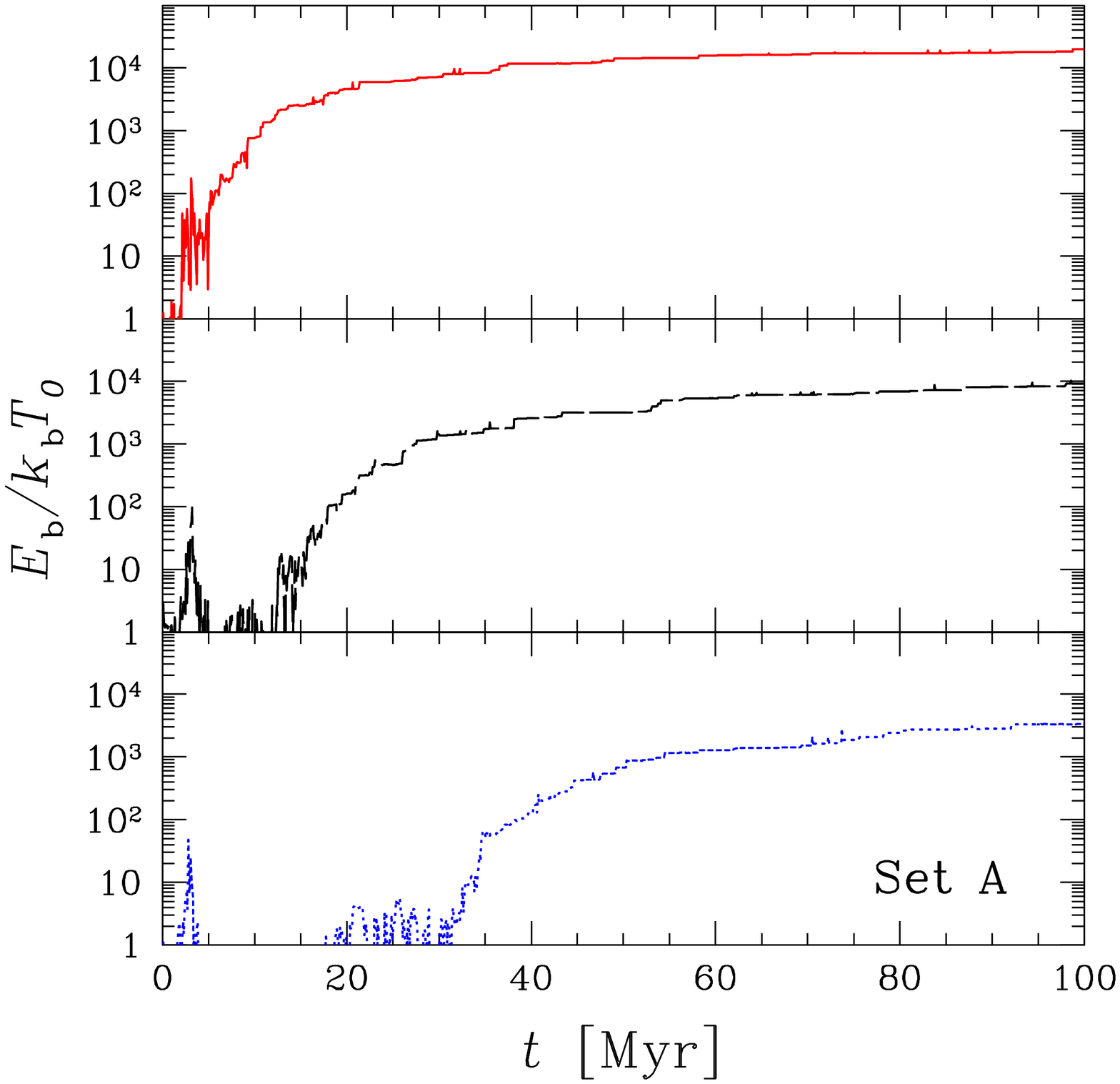,width=5.8cm} 
  \end{center}
  \caption{Core (left-hand panel) and half-mass (middle panel) radius as a function of time for the three considered metallicities. In the insets: zoom of the first 10 Myr. Right-hand panel: total internal energy of the binary content of the SCs as a function of time, normalised to the initial $k_{\rm b}T_0 = \frac{1}{3}\langle K \rangle|_{t=0}$, where $\langle K \rangle$ is the average kinetic energy of a star. Solid red line: $Z=0.01 \zsun$; dashed black line: $Z=0.1 \zsun$; dotted blue line: $Z=1 \zsun$. Each line is the median value of 10 simulated SCs with initial $r_{\rm vir}= 1 \pc$ and $W_0=5$ (set A).
}
\label{fig:3}
\end{figure*}

\begin{figure*}
  \begin{center}
    \epsfig{figure=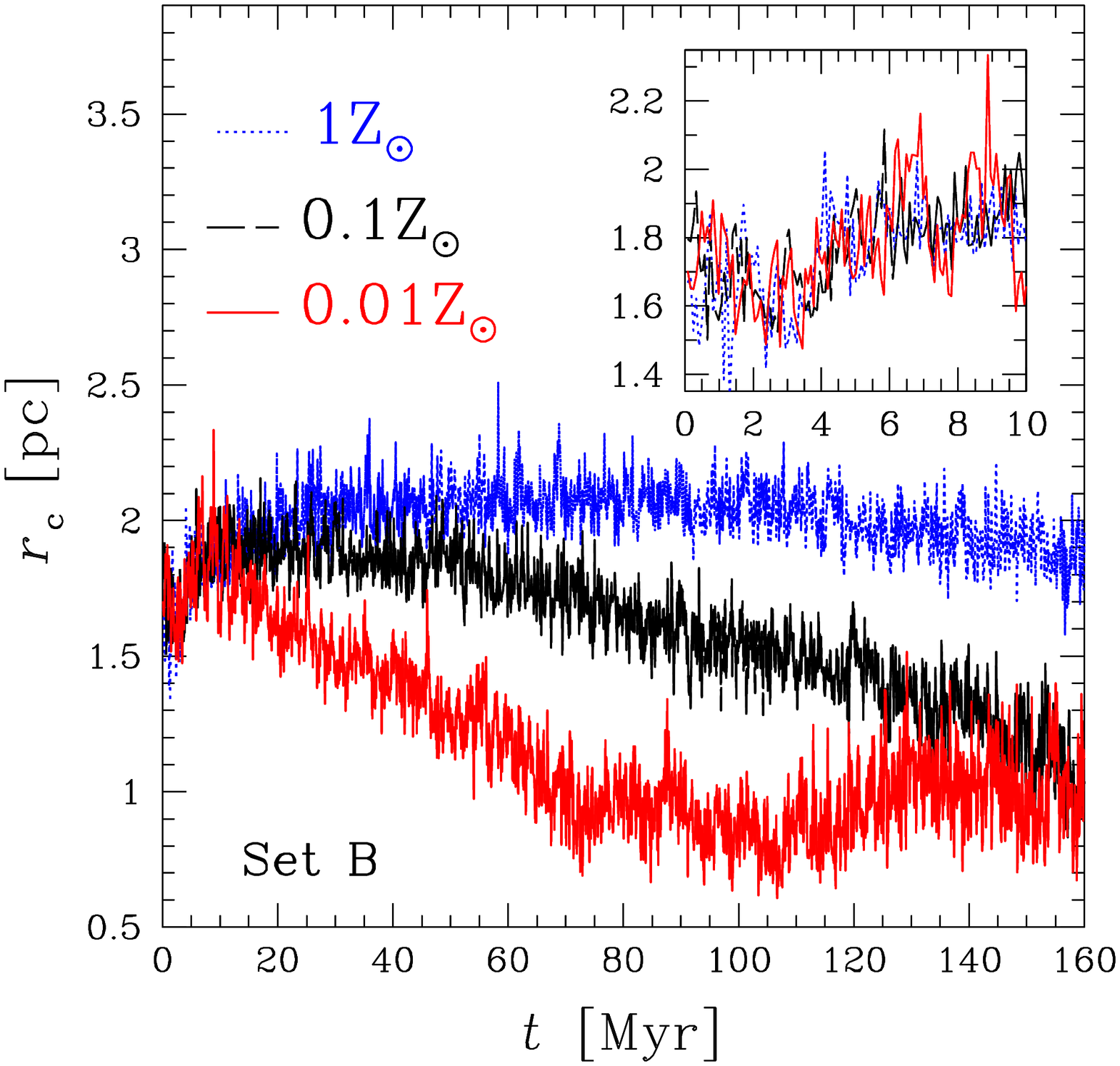,width=5.8cm} 
    \epsfig{figure=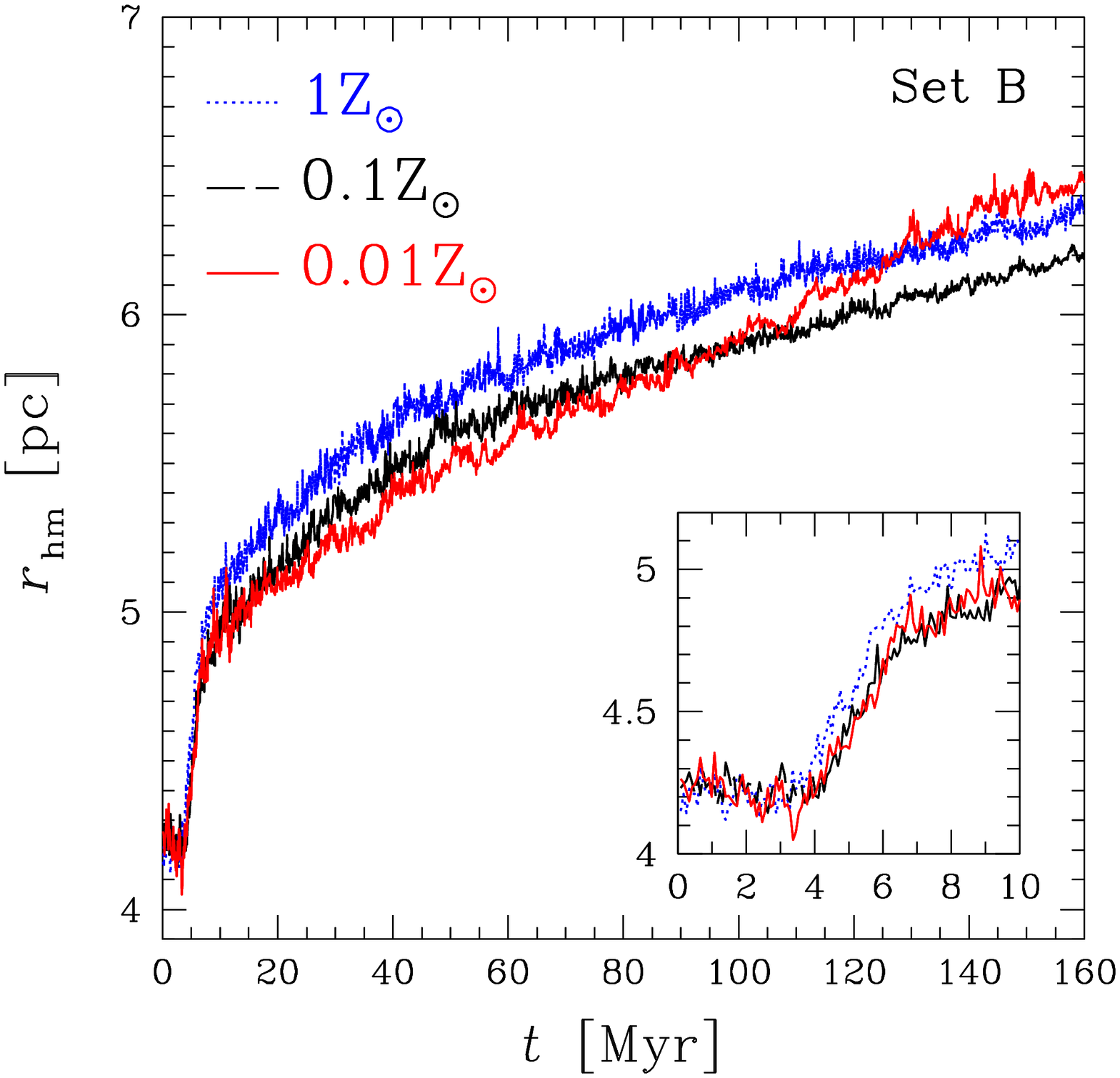,width=5.8cm} 
    \epsfig{figure=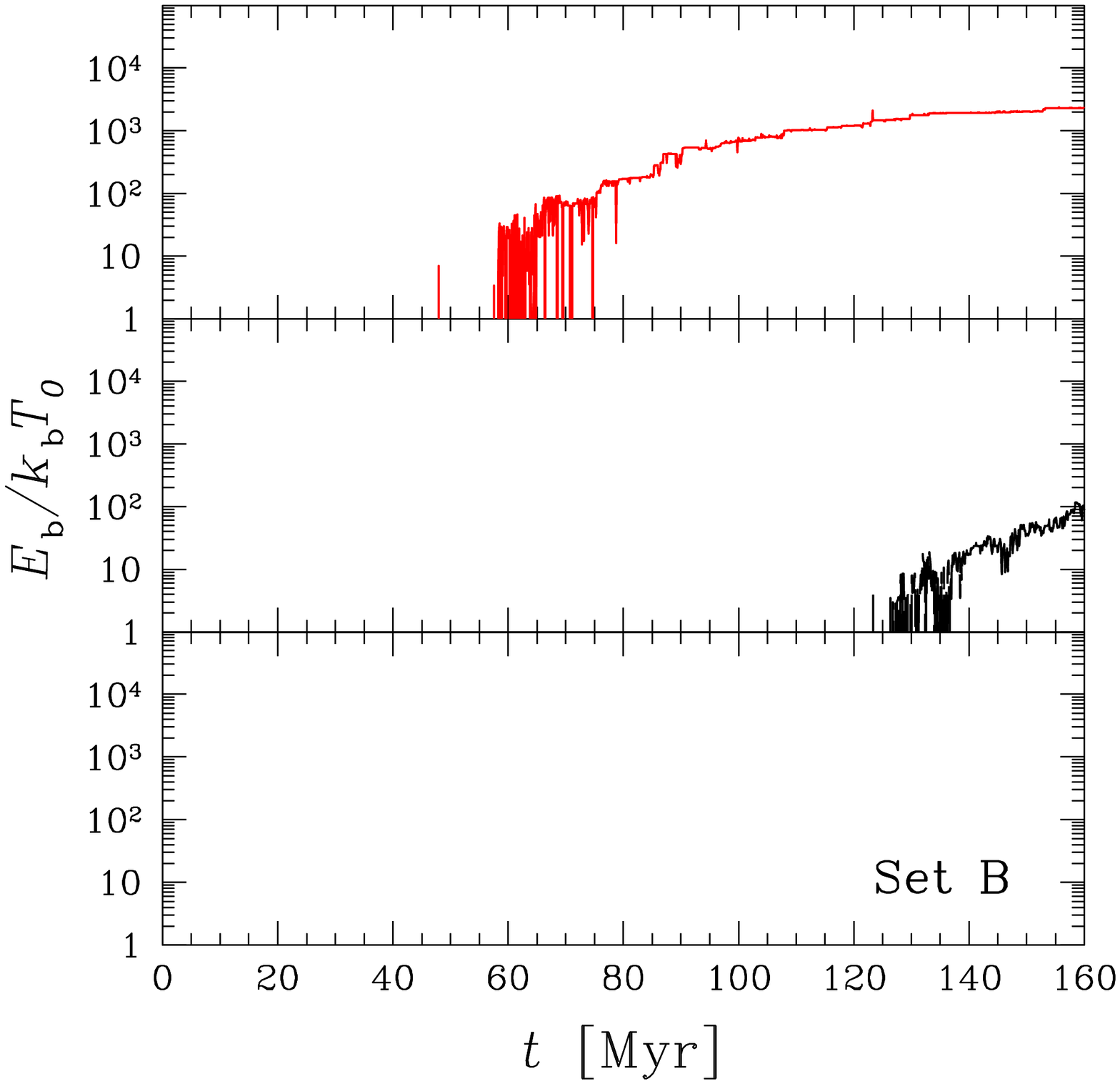,width=5.8cm} 
  \end{center}
  \caption{Same as Fig. \ref{fig:3}, but for simulations with $r_{\rm vir} = 5 \pc$ and $W_0=5$ (set B).
}
\label{fig:4}
\end{figure*}

\begin{figure*}
  \begin{center}
    \epsfig{figure=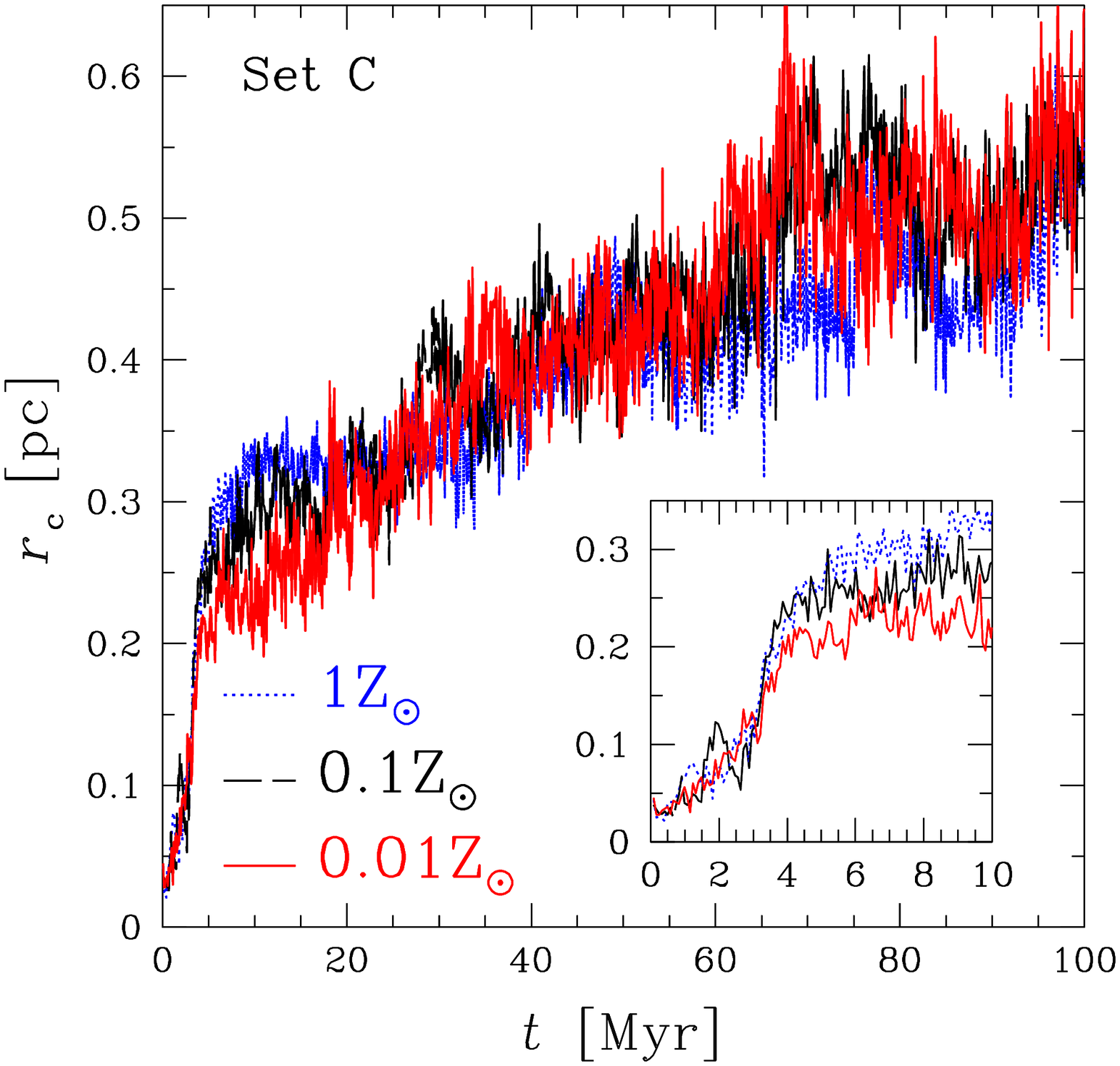,width=5.8cm} 
    \epsfig{figure=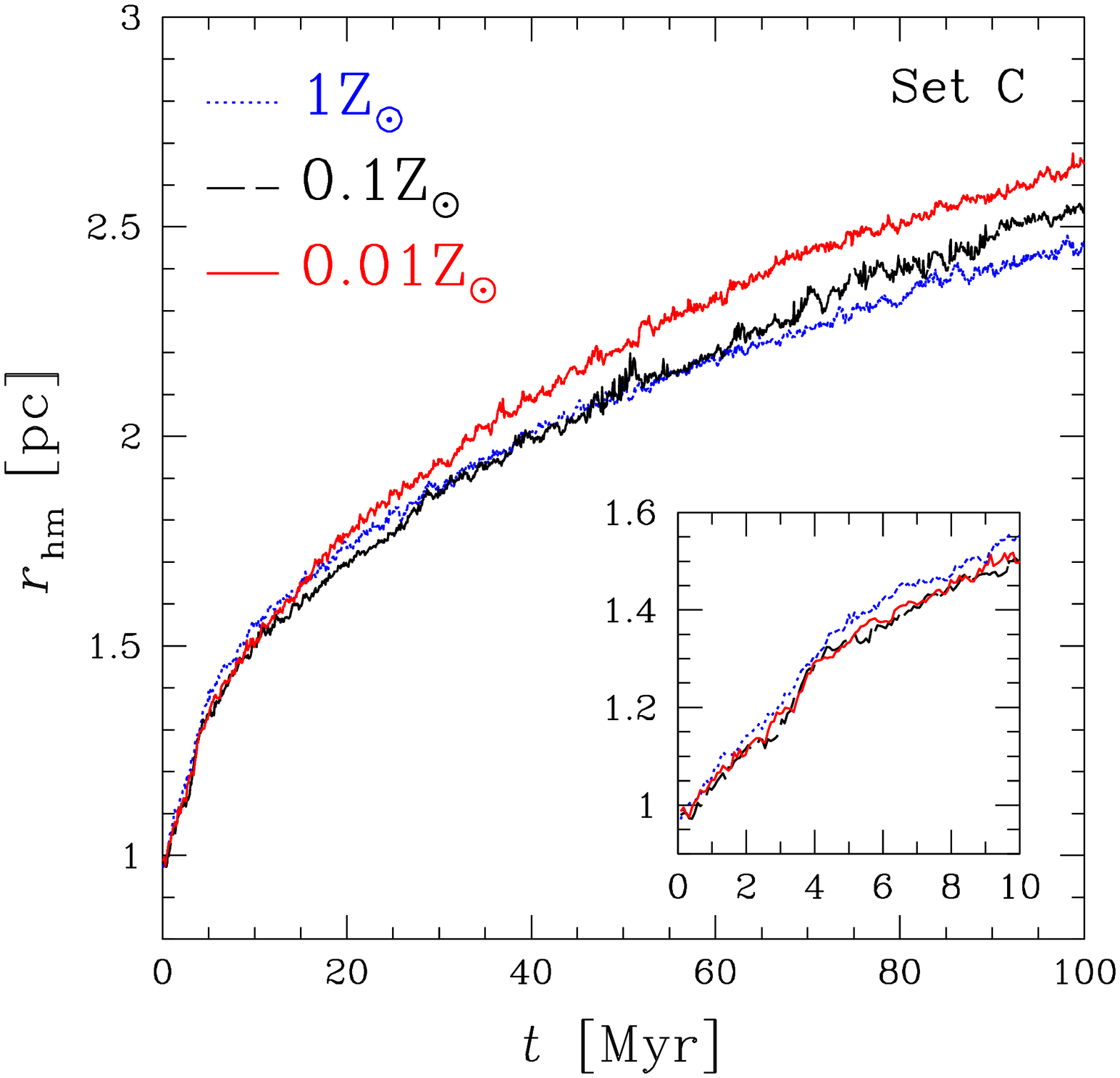,width=5.8cm} 
    \epsfig{figure=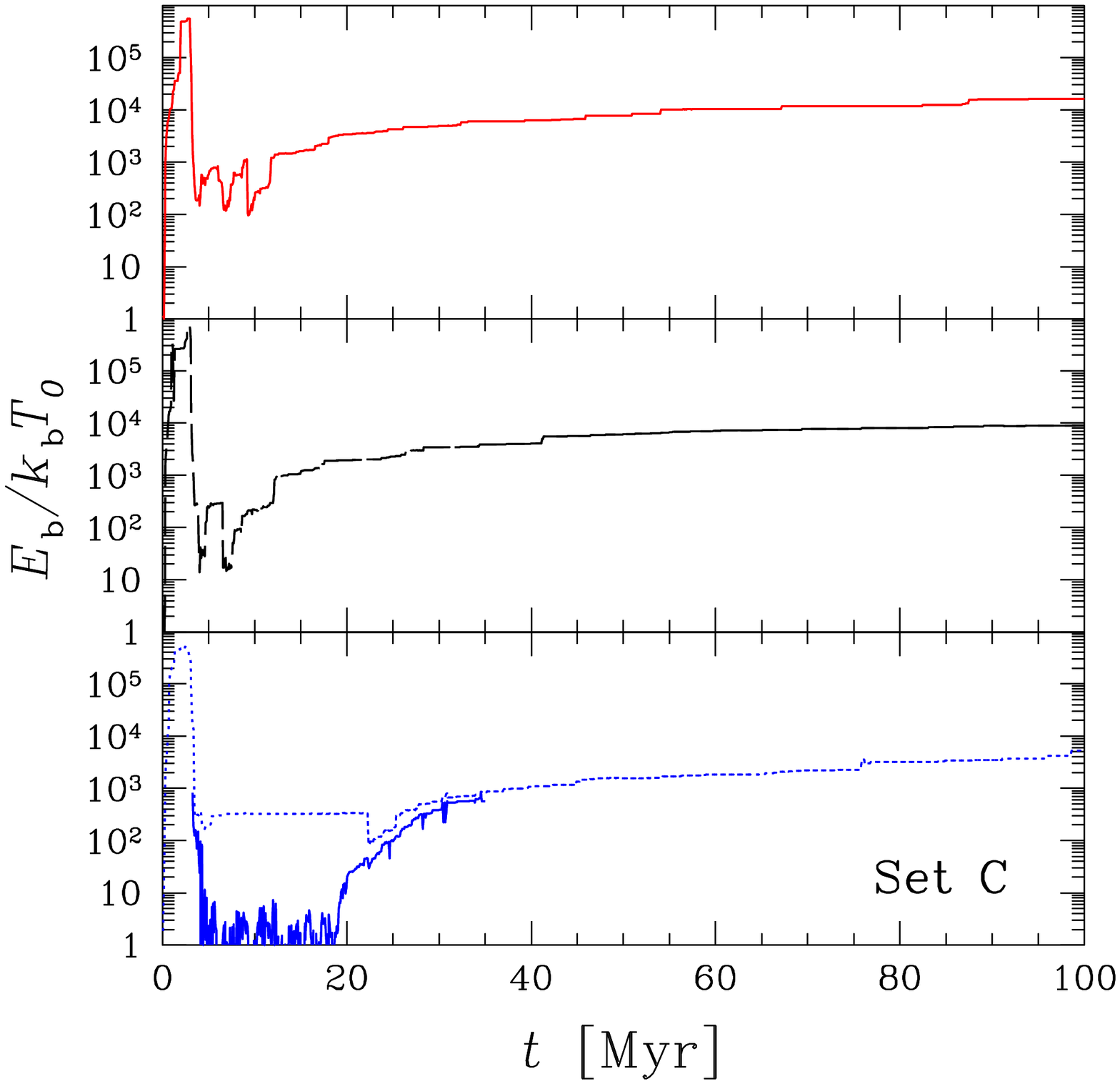,width=5.8cm} 
  \end{center}
  \caption{Same as Fig. \ref{fig:3}, but for simulations with $r_{\rm vir} = 1 \pc$ and $W_0=9$ (set C). The solid cyan line in the bottom right panel is the binding energy of the binaries in the core of the SCs with $Z=1 \zsun$. The plateau of the binary binding energy in $Z=1 \zsun$ SCs from 5 Myr to 30 Myr is due to an hard binary escaping from the core.
}
\label{fig:5}
\end{figure*}

We notice no further binary hardening after the initial expansion of the core in metal-rich SCs, at least until the core collapses again. The second collapse and the hardening of binaries varies from SC to SC, but it generally begins earlier at lower metallicity. The hardening of binaries begins around $20 \myr$ for the SCs with $Z=1 \zsun$ and at 10--15 Myr for the SCs with $Z=0.1 \zsun$ and $0.01 \zsun$. This behaviour is similar to the one observed in the SCs of set A, and indicates that stellar mass loss has sustained the initial expansion. Right-hand panel of Fig. \ref{fig:2} shows that two hard binaries are formed on average in each simulation. There are no significant differences in the number of binaries between SCs of different metallicity.

The middle panel of Fig. \ref{fig:5}, shows the evolution of the half-mass radius. We find that at 100 Myr, the half-mass radius of $Z=0.01 \zsun$ SCs is larger by 8 per cent than the one of $Z=1 \zsun$ SCs. 
The difference is less pronounced than in less concentrated SCs of set A, in agreement with the results of \citet{schul12}. Moreover, we notice that at 100 Myr the half-mass radius of the SCs with $W_0=9$ is larger by 30 per cent than the half-mass radius of the SCs with $W_0=5$.

\subsection{Core radius oscillations}
In the post-collapse phase, the core can still be subject to the gravothermal instability which has driven the collapse phase. In this case, the re-expansion of the core is quickly halted and the gravothermal catastrophe is restored. Then the core undergoes repeated contractions and re-expansions, which are called gravothermal oscillations.

Gravothermal oscillations were first discovered by \citet{bett&sugi84} by following the post-collapse phase of SCs using a gas model. The oscillations were later found also in Fokker-Plank calculations \citep{cohn89} and N-body simulations \citep{makino96}. These oscillations are called gravothermal, since gravothermal instability is thought to drive both expansion and contraction phases \citep{makino87,heggie94,makino96,mcmill96,breen&heggie12a,breen&heggie12b}.
While the collapse phase is always driven by the gravothermal instability, it is debated whether and under which conditions the expansion phase has a gravothermal nature. 

Most of our simulated SCs show core radius oscillations. These are not present in the core radius profile shown in Figs \ref{fig:3}, \ref{fig:4} and \ref{fig:5} since the oscillations cancel out when summing and averaging multiple SCs.

\begin{figure*}
  \begin{center}
    \epsfig{figure=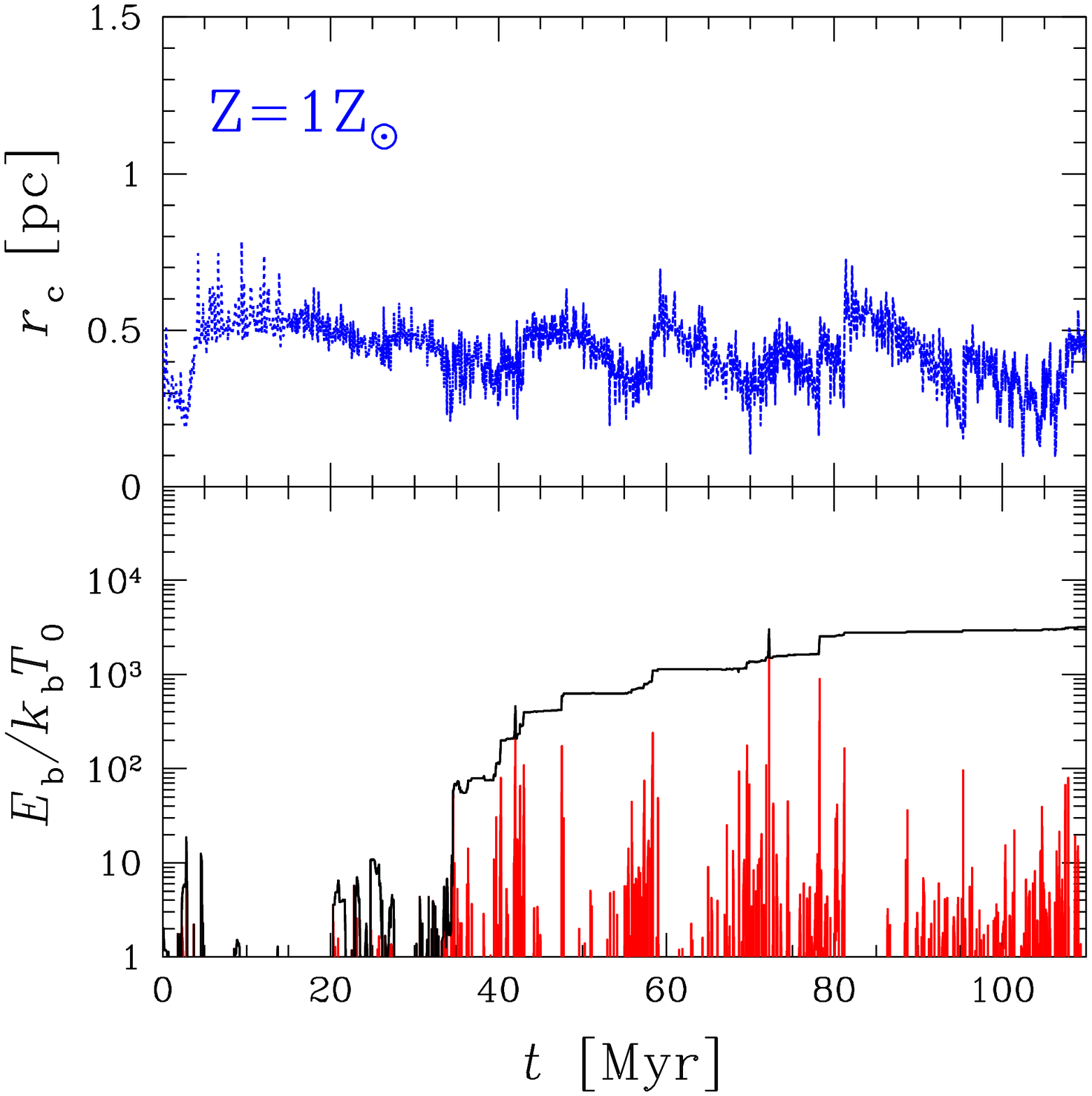,width=5.5cm} 
    \epsfig{figure=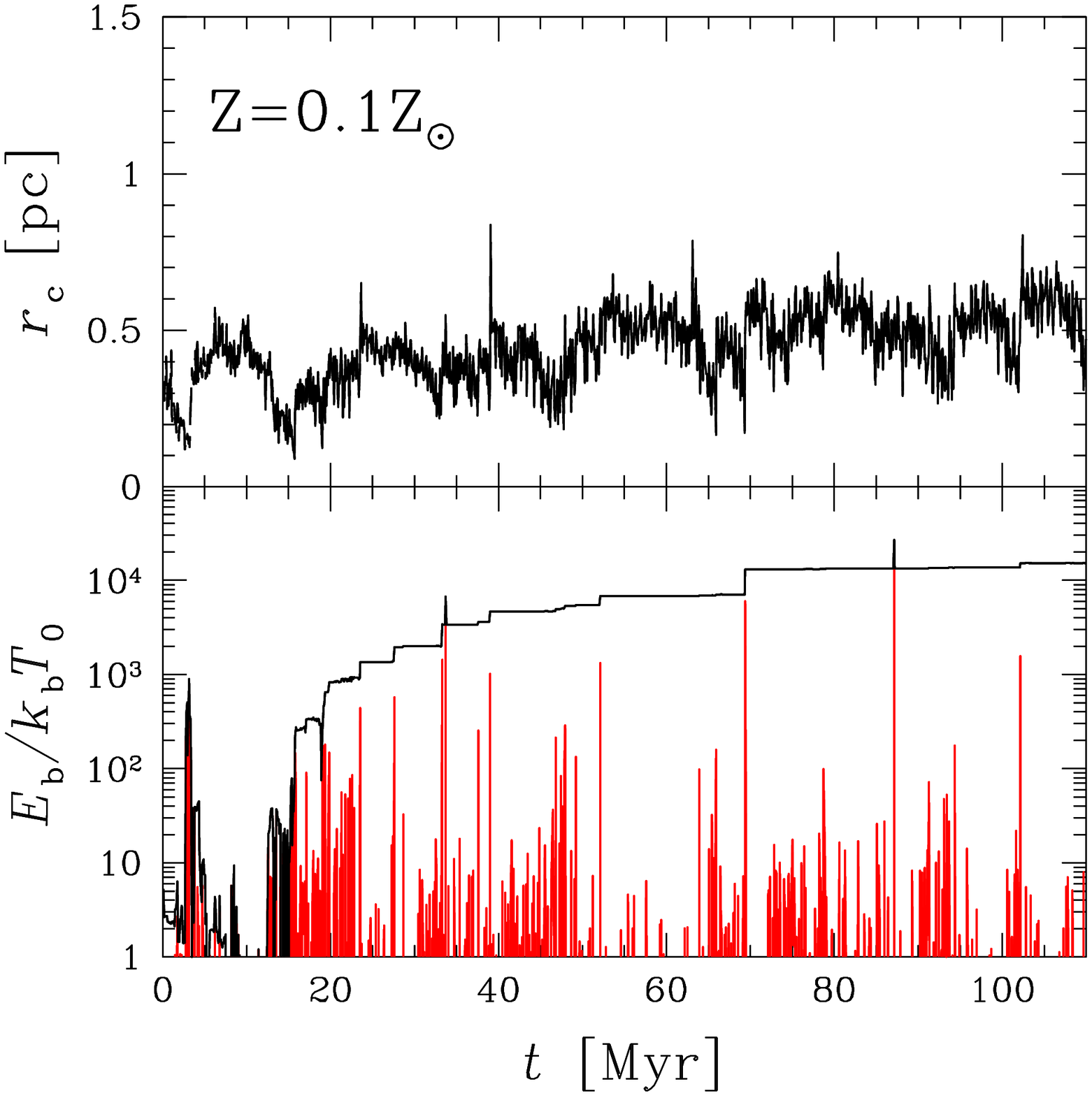,width=5.5cm} 
    \epsfig{figure=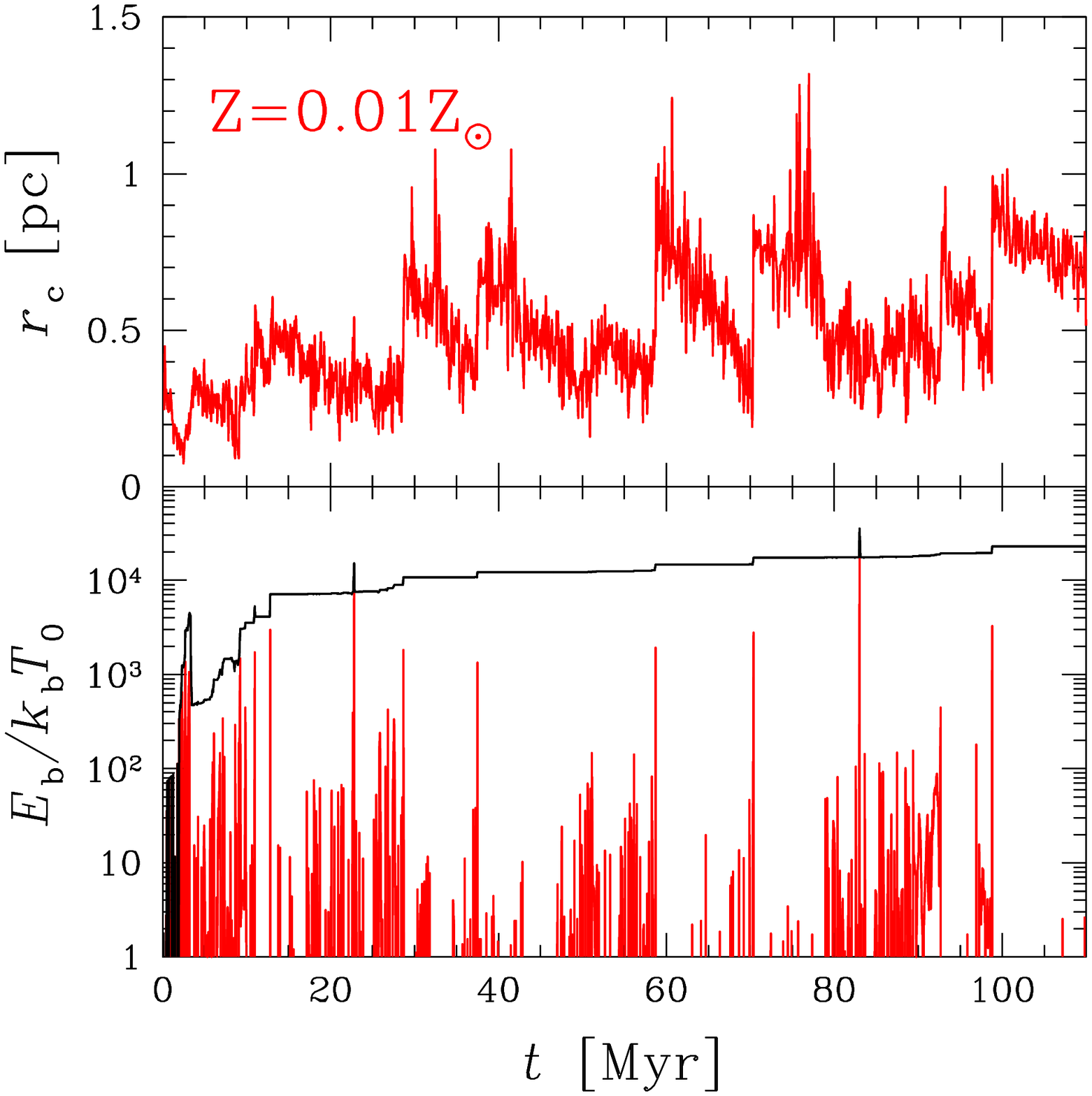,width=5.5cm} 
  \end{center}
  \caption{Top panels: core radius as a function of time for three selected clusters from Set A. Bottom panels: total internal energy of the binary content of the SCs as a function of time, normalised to the initial $k_{\rm}T_0 = \frac{1}{3}\langle K \rangle|_{t=0}$, where $\langle K \rangle$ is the average kinetic energy of a star. Black line: cumulative binary binding energy. Ochre line: increment in binary binding energy. Each line is obtained from single simulations with $r_{\rm vir}= 1 \pc$ and $W_0=5$ (set A). Left-hand panel: $Z=1 \zsun$. Middle panel: $Z=0.1 \zsun$. Right-hand panel: $Z=0.01 \zsun$.
}
\label{fig:6}
\end{figure*}

In Fig. \ref{fig:6}, we show the evolution of core radius and binary binding energy for three individual simulations of set A.
While the first core collapse occurs at 3 Myr for every realisation, the subsequent oscillations are stochastic and vary from SC to SC. After the first bounce at 3 Myr, the core collapse goes on and it is halted by a series of further core bounces. The resulting profile of the core radius versus time has a saw-tooth appearance.

The oscillations are metallicity dependent: number and amplitude of oscillations increase at lower metallicity. 
 From the lower panels of Fig. \ref{fig:6}, it is evident that every increase of core radius matches always the increments in the binary binding energy. This indicates that while the first bounce is mostly supported by mass loss by SNe and stellar winds, further bounces are supported only by binary hardening. The randomness of the oscillations is a consequence of the stochastic nature of three-body encounters.

The simulations of Fig. \ref{fig:6} represent the typical oscillations for the simulations of set A at the given metallicity. Individual simulations of set C present the same behaviour.
This is not the case for the simulations of set B, because some of them do not undergo a core collapse phase. In the case of set B, mild oscillations occur in one SC with $Z=0.1 \zsun$ and in most of the SC with $Z=0.01 \zsun$.

\section{Discussion}\label{sec:discussion}

\subsection{Interplay between dynamics and stellar evolution}\label{sec:disc1}

The expansion of the half-mass radius is the consequence of a heating mechanism being active in the core. This mechanism can be either binary hardening, 
or mass loss by stellar evolution. However, these two processes are metallicity dependent: stellar winds are inefficient at lower metallicity and do not contribute to reverse the core collapse. Thus, the core density increases more dramatically, and this enhances close encounters and thus binary hardening. A higher metallicity leads to stronger stellar-mass loss, which partially reverses core collapse, without strong binary hardening. Thus, we expect the half-mass radius of SCs with different metallicity to behave differently according to the dominant process that heats the SCs. If binary hardening is the dominant process, we expect the half-mass radius of metal-poor SCs to grow faster than that of metal-rich ones, because three-body encounters transfer more energy into the halo. In contrast, if stellar mass loss is the dominant process,  metal-rich SCs are expected to expand more with respect to metal-poor SCs, because stellar winds make the potential well shallower.

In both set A and set C, the half-mass radius of metal-poor SCs becomes larger than that of metal-rich ones, even if the size difference is more pronounced in set A than in set C. This indicates that binary hardening is responsible for the expansion of the SCs of both sets A and C. In contrast, metal-poor SCs of set B have a slightly smaller half-mass radius than that of metal-rich SCs until 80 Myr. After that time, the half-mass radius of SCs with $Z=0.01 \zsun$ begins to expand faster than that of metal-rich SCs. This means that, until the 80 Myr, the main heating mechanism of the SCs is stellar mass loss, but then binary hardening begins to be dominant in SCs with $Z=0.01 \zsun$.

The different evolution of the three simulated sets can be explained by considering how fast is the core collapse (expressed in terms of the core collapse timescale $t_{\rm cc}$) with respect to the lifetime of massive stars $t_{\rm se}\sim{}6$ Myr.

\subsection*{Set A}

SCs of set A have a half-mass relaxation timescale of $t_{\rm rh}\simeq 30 \myr$. If we assume $t_{\rm cc} \simeq 0.15$--$0.20$ $t_{\rm rh}$ (\citealt{port02}; \citealt{gurkan04}; see also \citealt{fujii} and Section \ref{sec:cctimes}) the core collapse proceeds simultaneously with the stellar mass loss ($t_{\rm cc} \sim t_{\rm se}$) in SCs of set A. In this situation, the interplay between the two processes is complicated.

Core collapse reaches its maximum at 3 Myr. While the first hard binaries begin to form, the first SNe remove mass from the SC and drive the expansion of the core. The expansion of the core is stronger in metal-rich SCs, because of the higher mass loss from the massive stars. The expansion also quenches three-body encounters. Only in the case with $Z=0.01 \zsun$  the core does not expand enough to quench the hardening, which instead goes on after the bounce.

The importance of stellar mass loss during the core bounce is confirmed by a set of test simulations without stellar evolution (see appendix \ref{chap:a1}). Though some hard binary form in the core collapse, the core bounce is mainly due to the mass loss by the SN explosions.

Fig. \ref{fig:3} shows that the binary binding energy at high metallicity ($Z\ge{}0.1 \zsun$) goes almost to zero during the expansion of the core. The main reason is that the binaries formed during the core collapse are unbound by the first SN explosions. In fact, the most massive stars are members of the first hard binaries, so that they are the first to undergo SN explosion. This is confirmed by the left-hand panel of Fig. \ref{fig:2} which shows that one hard binary is formed during the first $5\myr$, but then is disrupted in high metallicity ($Z\ge{}0.1 \zsun$) SCs.

As stellar mass loss becomes less intense, the core begins to recollapse. The recollapse is faster in metal-poor SCs than in metal-rich SCs, for two reasons: (i) the core of metal-poor SCs has become more dense and massive than that of metal-rich ones; (ii) metal-poor SCs have a higher maximum remnant mass, and core collapse in SCs with a mass spectrum tends to proceed on the dynamical friction timescale of the most massive stars, which shortens as the mass of the stars increases \citep{fujii}.

The reversal of the second core collapse can not be sustained by stellar mass loss, and eventually binary hardening is triggered. Because the second core collapse occurs faster in metal-poor SCs, binary hardening begins earlier (Fig. \ref{fig:3}). For this reason (and also for the higher frequency and strength of three-body encounters), more kinetic energy is extracted from binaries in metal-poor SCs than in metal-rich SCs. As soon as this kinetic energy is carried outwards by two-body relaxation, the rest of the SC expands and the half-mass radius increases accordingly. This results in a faster expansion of the half-mass radius in metal-poor SCs with respect to metal-rich ones. The stronger heating causes also the core radius of metal-poor SCs to become larger than the core radius of $Z=1 \zsun$ SCs. Overall, set A confirms the trend found by \citet{m&b13}, who simulate ten times less massive SCs.

\subsection*{Set B}
SCs of set B have a larger virial radius, which results in a longer half-mass relaxation timescale $t_{\rm rh} \simeq 308 \myr$. Thus, the core collapses when the most massive stars have already died for a long time ($t_{\rm cc} \gg t_{\rm se}$). In this regime, stellar evolution and dynamics are decoupled, and the evolution of the SCs  is characterized by an early stage dominated by stellar mass loss, followed by a late stage dominated by binary hardening. The stage dominated by stellar mass loss lasts longer for metal-rich SCs. 

After 60--70$\myr$ the half-mass radius of SCs with $Z=0.01 \zsun$ begins to expand faster than that of metal-rich SCs. We argue that this faster expansion is driven by the additional heating due to binary hardening. In fact, only $Z=0.01 \zsun$ SCs undergo significant heating by binary hardening before 100 Myr (Fig. \ref{fig:4}).

The differences in half-mass radius between SCs of different metallicity remain very small ($<10$ per cent) throughout the simulations. This can be due to the differences in stellar mass loss (Fig. \ref{fig:1}) being too small to produce significant differences in the SC expansion. However, we can not exclude that size differences due to binary hardening may become significant at $>160$ Myr, the time at which we stop the simulations.

\begin{figure}
  \begin{center}
    \epsfig{figure=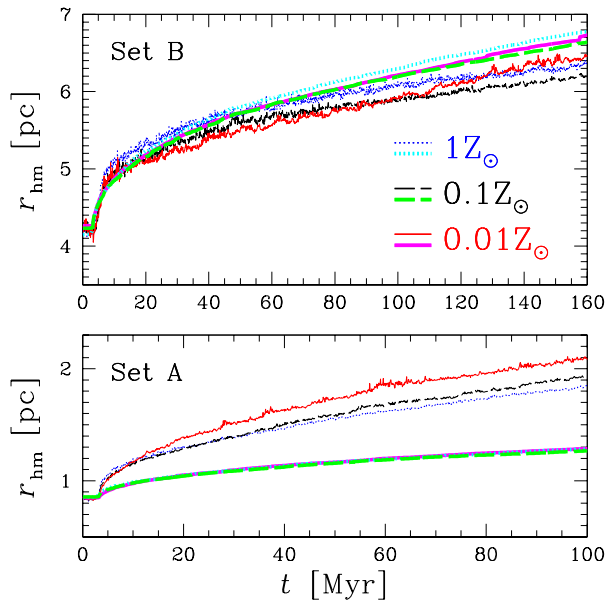,width=8.0cm} 
  \end{center}
  \caption{Half-mass radius as a function of time. The solid red thin line, the dashed black thin line and the dotted blue thin line are the median value of the half-mass radius obtained from the simulations with $Z=0.01 \zsun$, $Z=0.1 \zsun$ and $Z=1 \zsun$, respectively. 
The solid magenta thick line, the dashed green thick line and the dotted cyan thick line are computed using a semi-analytic prescription (see Section \ref{sec:disc1}) with $Z=0.01 \zsun$, $Z=0.1 \zsun$ and $Z=1 \zsun$, respectively. The lines representing the semi-analytic prescription are smoother than the lines of the N-body simulations. In the bottom panel, the lines of the semi-analytic prescription overlap. Top panel: SCs with initial $r_{\rm vir}=5 \pc$ and $W_0=5$ (set B). Bottom panel: SCs with initial $r_{\rm vir}= 1 \pc$ and $W_0=5$ (set A).
}
\label{fig:7}
\end{figure}

In the case of the simulations of set B, the stellar mass loss is the source of energy initially driving the expansion of the SCs. To check this, we compare  the time evolution of the half-mass radius in the simulations with that of an analytic model (Fig. \ref{fig:7}). 
In this model, we assume that the half-mass radius changes according to the expansion of the SC due to mass loss. Then, mass is ejected in a time shorter than the dynamical time, the half-mass radius $r_{\rm hm}$ is related to the total mass $M$ of the SC by the formula \citep[see e.g.][]{hills80}:
\begin{equation}
\frac{r_{\rm hm}(t)}{r_{\rm hm}(0)} = \frac{M(t)/M(0)}{2 M(t)/M(0) - 1}
\end{equation}

For $M(t)$, $M(0)$ and $r_{\rm hm}(0)$ we use the values obtained from the simulations. In particular, for $M(t)$ we use the total bound mass of the SCs, to take into account the escapers due to SN kicks.

This impulsive approximation is  valid for the first SNe, but begins to overestimate the half-mass radius expansion at later times, when the mass loss rate slows down. Nonetheless, Fig. \ref{fig:7} fairly reproduces the evolution of the half-mass radius for set B in the first $\sim{}60$ Myr. 
This semi-analytic prescription fails to reproduce the evolution of the half-mass radius for set A, because the expansion is mostly due to binary hardening.

\subsection*{Set C}
SCs of set C have the same size and mass as SCs of set A, but are much more concentrated. They have a core relaxation timescale shorter than $1 \myr$, which means that the core is already collapsed at the beginning of the simulations ($t_{\rm cc} \ll t_{\rm se}$). The initial core density of these SCs is high enough to make binary hardening the dominant process involved in the reversal of the core collapse.

As in the SCs of set A, the energy generated by three-body encounters is higher in metal-poor SCs. As a result, metal-poor SCs expand more than metal-rich SCs. However, the size differences between SCs of different metallicity are smaller with respect to the set A (8 per cent, compared to 14 per cent).
We argue that stellar mass loss contributes to the expansion to a lesser extent with respect to SCs of set A. Stellar mass loss is less important than dynamics and the differences arising from different metallicities are less evident. In fact, SCs of set C expand more than SCs of set A, and this is due to the stronger heating by binary hardening.
 

\subsection{Core oscillations: gravothermal or not?}

We investigated whether or not the oscillations found in the simulations of set A are gravothermal. While the contracting phase of the oscillations is always driven by the gravothermal instability, the expanding phase may not. According to \citet{mcmill96}, the evidence of gravothermal behaviour is a prolonged expansion of the core, during which there is no binary heating. This would mean that the energy required to expand the core is flowing from outside the core, rather than being generated by three-body encounters in the core.

We do not find any significant prolonged expansion of the core without binary activity. We argue that most of the expansion phases of the oscillations are not gravothermal, but are driven only by binary hardening. Another clue of the non-gravothermal behaviour of the oscillations is that many core bounces are very rapid. This is mostly evident in the SCs with $Z=0.01 \zsun$ (Fig. \ref{fig:6}, lower panel), in which the increase in core radius is very discontinuous. Gravothermal oscillations should exhibit a much longer expansion phase, which proceeds on the relaxation timescale of the core \citep{bett&sugi84}. Furthermore, after that the rapid expansion occurs, the core immediately begins to lose kinetic energy, i.e. the collapse driven by gravothermal instability is immediately restored. If an energy flux was established from the inner halo to the core, the core would not lose kinetic energy so quickly and the transition between expansion and contraction phase would be more gradual. 
The increase of oscillations at lower metallicity confirms that these oscillations are related to strong three-body encounters, rather than to an inverse temperature gradient.

\begin{figure}
  \begin{center}
    \epsfig{figure=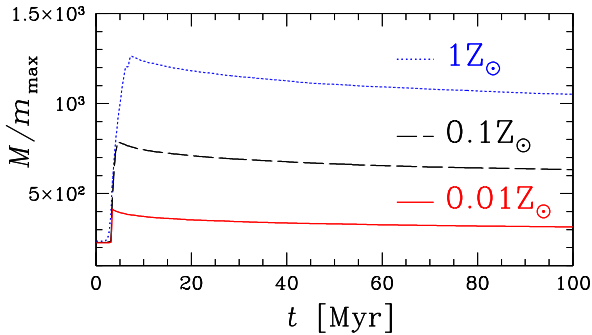,width=8.0cm} 
  \end{center}
  \caption{Ratio of total mass of the SC $M$ and the maximum stellar mass $m_{\rm max}$ as a function of time for our simulated SCs. Solid red line: $Z=0.01 \zsun$; dashed black line: $Z=0.1 \zsun$; dotted blue line: $Z=1 \zsun$.
}
\label{fig:8}
\end{figure}

This result is consistent with the criterion of \citet{breen&heggie12b} for gravothermal oscillations in multi-mass systems. They argued that gravothermal oscillations should occur for SCs in which $N_{\rm eff}=M/m_{\rm max}\gtrsim 10^4$. As shown in Fig. \ref{fig:8}, $N_{\rm eff}$ is always well below this limit, especially for the metal-poor SCs.

In particular, the expanding phase of the oscillations in SCs with $Z=0.01 \zsun$ are very rapid and dramatic. They correspond to a sudden increase of binary binding energy due to a single, strong three-body encounter occurred in the core. These three-body encounters are associated with the ejection of a binary or of a massive star.

\subsection{Core collapse time}\label{sec:cctimes}

The SCs of set A have a relaxation timescale of $t_{\rm rh} \simeq 36 \myr$. If we assume $t_{\rm cc} \simeq 0.2 t_{\rm rh}$ \citep{port02}, we expect the core collapse to take place at $t_{\rm cc}\simeq 7 \myr$. However, we find that the core collapse occurs much earlier, at $t_{\rm cc}\simeq 3 \myr$. In particular, the SCs of set A undergo core collapse at the same time as the SCs simulated by \citet{m&b13}, which have $N=5000$ and a half-mass relaxation time of $t_{\rm rh}\simeq 12 \myr$. 

The most likely explanation is provided by \citet{fujii}. They find that the core collapse time scales as $t_{\rm cc}/t_{\rm rc}\propto (m_{\rm max} / \langle m \rangle)^{-1}$, where $\langle m \rangle$ is the mean stellar mass and $m_{\rm max}$ the upper mass limit of the IMF. This relation starts to deviate at a larger values of $m_{\rm max} / \langle m \rangle$ for models with low $N$ \citep[see figure 6 of][]{fujii}. In particular, \citet{fujii} find that this scaling breaks for $N_{\rm eff}=M/m_{\rm max} \lesssim 100$, below which the system starts to behave chaotically.

While the SCs of set A and those in \citet{m&b13} have the same ratio $m_{\rm max} / \langle m \rangle \simeq 230$, the former have $N_{\rm eff}\simeq 216$ while the latter have $N_{\rm eff}\simeq 23$. This means that the ratio $t_{\rm cc}/t_{\rm rc}$ is different for each SC model. In fact, the left-hand panel of figure 6 in \citet{fujii} shows that the ratio $t_{\rm cc}/t_{\rm rc}$ of the two models with $m_{\rm max}/\langle m \rangle \simeq 258$, $N=2k$ and $N=32k$ differs roughly by a factor of $1/3$. Since the $t_{\rm rh}$ of the SCs of set A is 3 times the $t_{\rm rh}$ of the SCs in \citet{m&b13}, the factor $1/3$ cancels the differences in the relaxation timescale and leads to the same $t_{\rm cc}$ for both simulations.

\section{Conclusions}\label{sec:conclusions}

We ran direct N-body simulations to investigate the impact of stellar evolution and dynamics on the structural properties of SCs. Three sets of initial conditions were used to vary the core relaxation timescale $t_{\rm rc}$ of the SCs and thus the importance of dynamics. We expect that the efficiency of stellar evolution with respect to dynamical heating depends on the ratio between the core collapse timescale $t_{\rm cc}$ and the lifetime of massive stars $t_{\rm se}$. We consider three cases: $t_{\rm cc} \sim t_{\rm se}$, $t_{\rm cc} \gg t_{\rm se}$, $t_{\rm cc} \ll t_{\rm se}$.

Metallicity plays an important role: metal-rich SCs lose more mass than metal-poor SCs because of stellar winds and SNe. We find that the main effect of stellar mass loss is to delay the hardening of binaries, and this delay is more severe for higher metallicity. 


We found size differences in SCs with different metallicity. The differences are more significant in the simulations with $t_{\rm cc} \sim t_{\rm se}$ (set A), for which we find that at $t = 100 \myr$ metal-poor SCs have a 14 per cent larger half-mass radius than metal-rich SCs. Similar size differences were found in the simulations with $t_{\rm cc} \ll t_{\rm se}$ (set C). Simulations with $t_{\rm cc} \gg t_{\rm se}$ (set B) do not show significant size differences between SCs of different metallicity. 

This result can be explained as follows. 
\begin{itemize}
\item  In the SCs of set A ($t_{\rm cc} \sim t_{\rm se}$), stellar mass loss contributes to the reversal of core collapse, and this contribution becomes more important at higher metallicity. As a result, the hardening of binaries begins later for metal-rich SCs, even if the core collapse occurs at the same time regardless of the metallicity. The expansion of the core after collapse (i.e. the core bounce) is larger for high metallicity. 

Thus, the core of metal-poor SCs remains much denser, and the hardening of binaries begins earlier at low metallicity. Moreover, the hardening of binaries is enhanced by the more massive remnants of metal-poor SCs. Due to the enhanced heating, the half-mass radius of metal-poor SCs expands faster and, by the end of the simulation, it is 14 per cent larger than the half-mass radius of metal-rich SCs

The SCs of set A exhibit the same behaviour as the less massive SCs simulated by \citet{m&b13}, except for the core radius oscillations. In particular, the size differences between SCs with different metallicity are consistent with the results of \citet{m&b13} and \citet{schul12}.

\item  For SCs of set B ($t_{\rm cc} \gg t_{\rm se}$), the delay of binary hardening is a consequence of the delayed core collapse. During the first $\sim{} 60 \myr$, the evolution of the SCs is ruled by stellar evolution. During this time, the SCs experience an expansion of the core, which lasts longer for higher metallicity. Stellar mass loss also drives the expansion of the half-mass radius. Before the core collapse of the metal-poor SCs, the half-mass radius of metal-rich SCs is 5 per cent larger than that that of metal-poor SCs. 

Once metal-poor SCs experience core collapse, the injection of energy by three-body encounters begins in their core. As a consequence, metal-poor SCs begin to expand more than metal-rich ones, and the half-mass radius of metal-poor SCs becomes larger than the half-mass radius of metal-rich SCs. However, on average, the differences in half-mass radius between SCs with different metallicity remain $<10$ per cent, throughout the simulations.

\item  For SCs of set C ($t_{\rm cc} \ll t_{\rm se}$), the core of the SCs is already collapsed at the beginning of the simulations. The initial core density of these SCs is so high  that binary hardening is the dominant process involved in the reversal of the core collapse. Since stellar mass loss is less important than dynamics, the differences arising from different metallicities are less evident. After the reversal of core collapse, the evolution of the simulated SCs with $W_0=9$ is qualitatively similar to the evolution of SCs with $W_0=5$, but the size differences between SCs with different metallicity are smaller.
\end{itemize}

Finally, we found core radius oscillations in the simulated SCs with $r_{\rm vir}=1 \pc$ and in some of the SCs with $r_{\rm vir}=5 \pc$. These oscillations grow in number and amplitude as metallicity decreases. We investigated whether the expansion phase of these oscillations was driven by gravothermal instability or by strong three-body interactions occurring in the core. We concluded that most of the oscillations are not gravothermal, but they are associated with the ejection of massive stars and binaries from the core.

In summary, we confirm that the interplay between metallicity-dependent stellar evolution and dynamical heating is a crucial ingredient to understand the evolution of young SCs. In forthcoming studies, we will investigate how the physics of gas evaporation and the presence of strong tidal fields can affect this scenario.

\section*{Acknowledgements}
We thank the referee, Peter Anders, for his careful reading of the manuscript and for his useful comments. We would like to thank Monica Colpi, L\'eo Girardi, Stefano Rubele and Emanuele Ripamonti for useful discussions. We acknowledge the CINECA Awards N. HP10CGUBV0 and HP10C894X7 for the availability of high performance computing resources and support. We made use of the public software package {\sc Starlab} and of the {\sc SAPPORO} library \citep{sapporo} to run Starlab on graphics processing units (GPUs). We acknowledge all the developers of Starlab, and especially its primary authors: Piet Hut, Steve McMillan, Jun Makino, and Simon Portegies Zwart. We thank the authors of SAPPORO, and in particular E. Gaburov, S. Harfst and S. Portegies Zwart.
 MM and AAT acknowledge financial support from the Italian Ministry of Education,
University and Research (MIUR) through grant FIRB 2012 RBFR12PM1F. MM acknowledges financial support from INAF through grant PRIN-2011-1 and from CONACyT through grant 169554.

\begin{appendix}
\section{Runs without stellar mass loss}\label{chap:a1}
\begin{figure*}
  \begin{center}
    \epsfig{figure=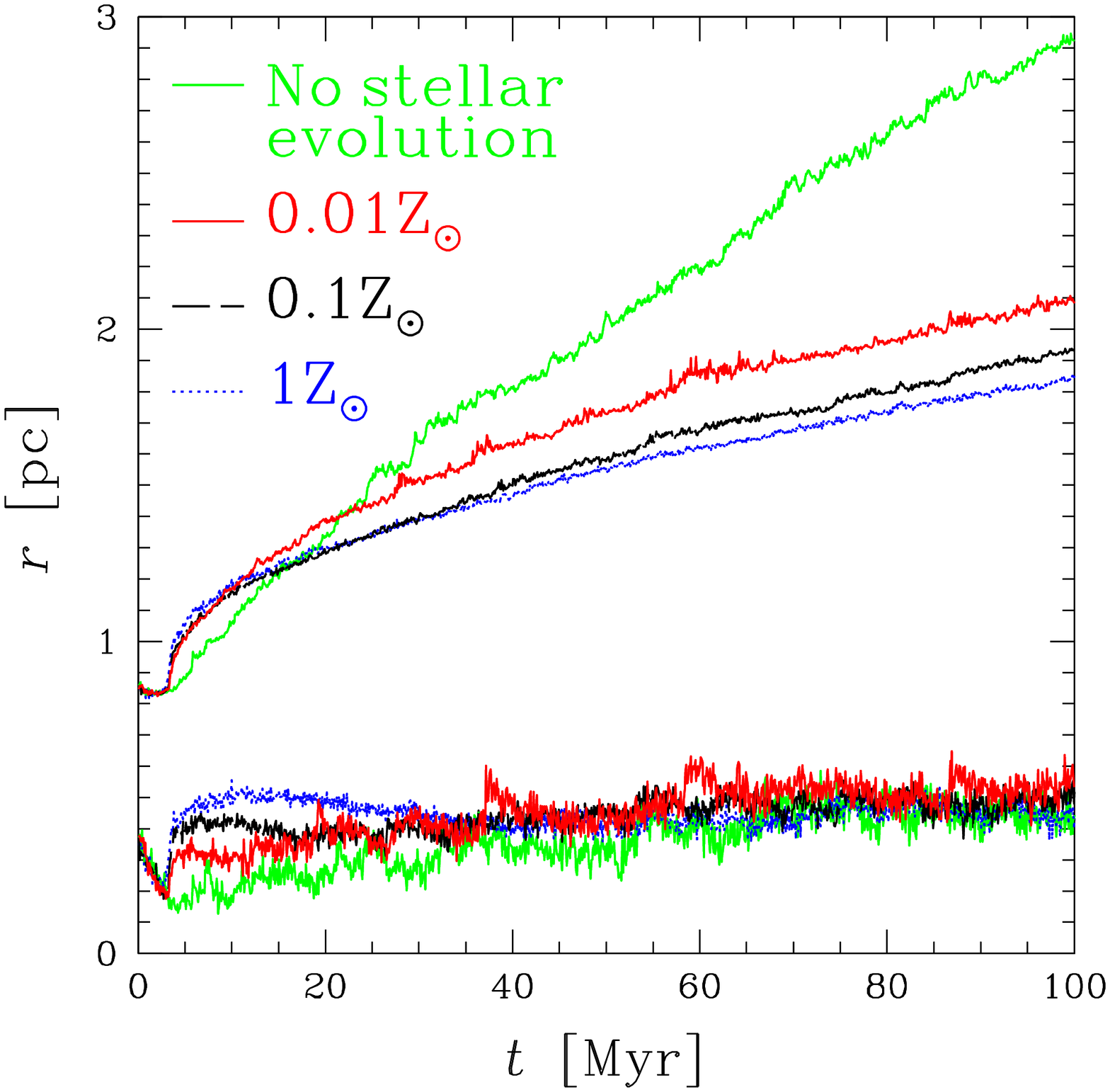,width=8.0cm} 
    \epsfig{figure=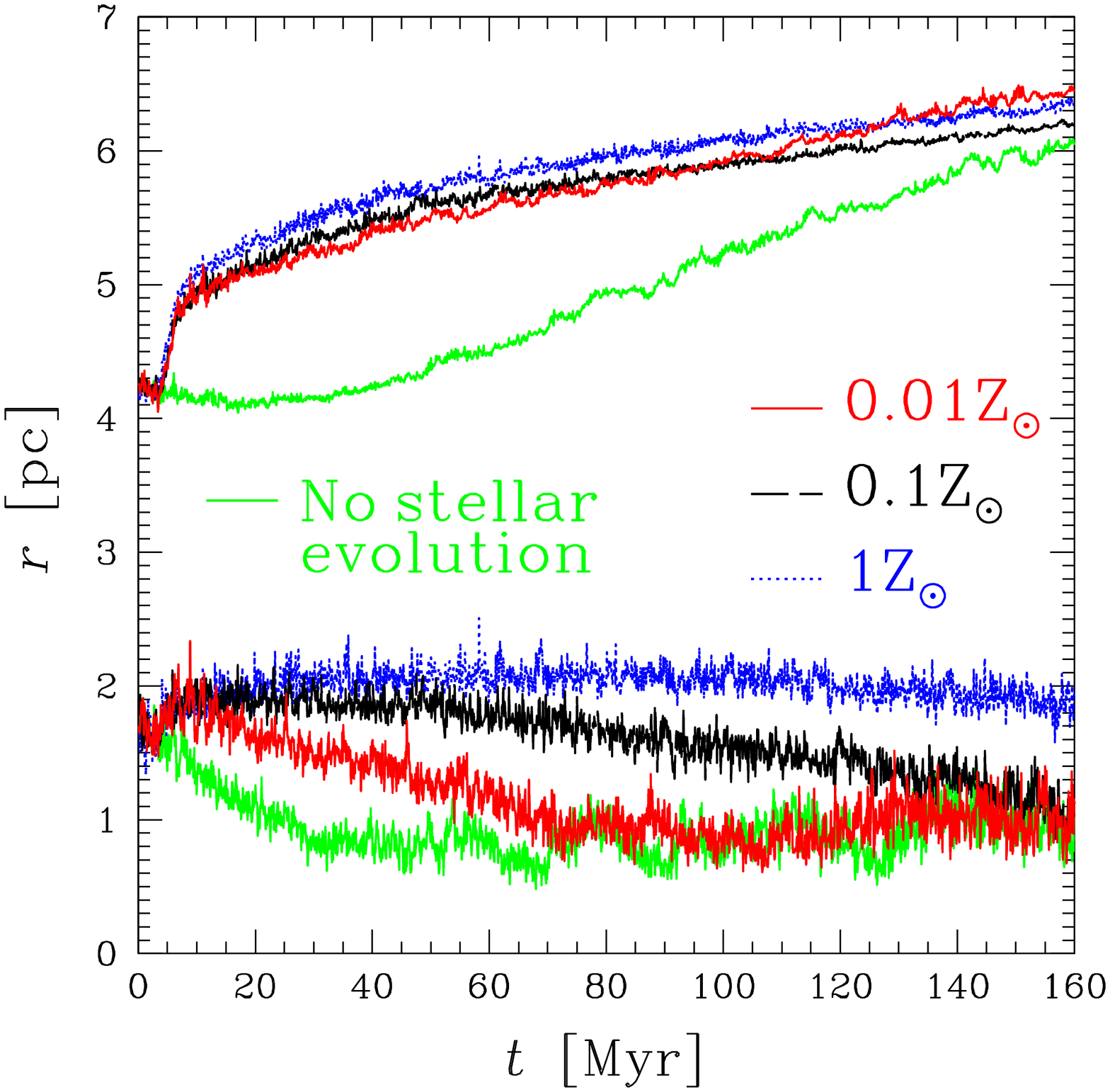,width=8.0cm} 
  \end{center}
  \caption{Core (bottom lines) and half-mass (top lines) radius as a function of time for the three considered metallicities, and for simulations without stellar evolution. Solid red line: $Z=0.01 \zsun$; dashed black line: $Z=0.1 \zsun$; dotted blue line: $Z=1 \zsun$. Solid green line: no stellar evolution. Left-hand panel: SCs with $r_{\rm vir}= 1 \pc$ and $W_0=5$ (set A); right-hand panel: SCs with $r_{\rm vir}= 5 \pc$ and $W_0=5$ (Set B). Each line is the median value of 10 simulated SCs.}
\label{fig:9}
\end{figure*}

We simulated two additional sets (of 10 SCs each) with the same initial conditions as set A and B, but without stellar evolution. We switched off stellar winds and SN explosions, so that stars do not lose mass throughout simulations. The simulations without stellar evolution are the extreme case in which dynamical heating is the only process driving the expansion of the SCs. We plot the evolution of core and half-mass radius of these simulations in Fig. \ref{fig:9}, in comparison with that of the runs with stellar evolution. 

The left-hand panel of Fig. \ref{fig:9} confirms that stellar mass loss does not influence the core collapse time. Comparing the core radii, it is apparent that the initial expansion of the core is caused only by stellar evolution. Also the expansion of the half-mass radius is strongly influenced by stellar evolution. The increase in half-mass radius at 3 Myr is stronger in SCs with stellar evolution, since it is mainly due to the first SN explosions. However, at late times, the half-mass radius in the simulations without stellar evolution increases much faster than that of SCs with stellar evolution, due to the enhanced dynamical heating.

In the right-hand panel of Fig. \ref{fig:9} it is evident that the weak core collapse in the first 3 Myr of SCs with of set B is the beginning of a longer core collapse. In the simulations with stellar evolution, this collapse is interrupted by the sudden ejection of mass by the first SNe. From the comparison of the half-mass radius, it is also clear that the increase of the half-mass radius after 3 Myr is only due to the stellar evolution.

\end{appendix}

\begin{thebibliography}{99}

\bibitem[Angeletti \& Giannone(1977)]{a&g77}Angeletti L., Giannone P., 1977, A\&A, 58, 363
\bibitem[Angeletti \& Giannone(1980)]{a&g80}Angeletti L., Giannone P., 1980, A\&A, 85, 113
\bibitem[Applegate(1986)]{applegate86}Applegate J.H., 1986, ApJ, 301, 132
\bibitem[Belczynski et al.(2010)]{b10}Belczynski K., Bulik T., Fryer C.L., Ruiter A., Valsecchi F., Vink J.S., Hurley J.R., 2010, ApJ, 714, 1217
\bibitem[Belkus et al.(2007)]{belkus07}Belkus H., Van Bever J., Vanbeveren D., 2007, ApJ, 659, 1576
\bibitem[Bettwieser \& Sugimoto(1984)]{bett&sugi84}Bettwieser E., Sugimoto D., 1984, MNRAS, 204, 493
\bibitem[Binney \& Tremaine(2008)]{bin&tre}Binney J., Tremaine S., 2008, Galactic dynamics: Second Edition, Princeton University Press
\bibitem[Breen \& Heggie(2012a)]{breen&heggie12a}Breen P.G., Heggie D.C., 2012a, MNRAS, 420, 309
\bibitem[Breen \& Heggie(2012b)]{breen&heggie12b}Breen P.G., Heggie D.C., 2012b, MNRAS, 425, 2493
\bibitem[Chernoff \& Weinberg(1990)]{chernoff&90}Chernoff D.F., Weinberg M.D., 1990, ApJ, 351, 121
\bibitem[Cohn et al.(1989)]{cohn89}Cohn H., Hut P., Wise M., 1989, ApJ, 342, 814
\bibitem[Downing(2012)]{down12}Downing J.M.B., 2012, MNRAS, 425, 2234
\bibitem[Elson, Hut, \& Inagaki(1987)]{elson87}Elson R., Hut P., Inagaki S., 1987, ARA\&A, 25, 565
\bibitem[Fryer \& Kalogera(2001)]{fryer01}Fryer C.L., Kalogera V., 2001, ApJ, 554, 548
\bibitem[Fryer et al.(2012)]{fryer12}Fryer Ch. L., Belczynski K., Wiktorowicz G., Dominik M., Kalogera V., Holz D. E., 2012, ApJ, 749, 91
\bibitem[Fryer(1999)]{fryer99}Fryer C.L., 1999, ApJ, 522, 413
\bibitem[Fujii \& Portegies Zwart(2014)]{fujii}Fujii M.S., Portegies Zwart S., 2014, MNRAS, 439, 1003
\bibitem[Gaburov, Harfst \& Portegies Zwart(2009)]{sapporo}Gaburov E., Harfst S., Portegies Zwart S., 2009, NewA, 14, 630
\bibitem[Gieles(2013)]{gieles13}Gieles M., 2013, ASP Conf. Ser. Vol. 470, San Francisco, Astron. Soc. Pac., 339
\bibitem[Gonz\'alez Delgado et al.(2003)]{gonzalez03}Gonz\'alez Delgado D., Olofsson H., Kerschbaum F., Sch\"oier F. L., Lindqvist M., Groenewegen M. A. T., 2003, A\&{}A, 411, 123
\bibitem[Goodman(1987)]{goodman87}Goodman J., 1987, ApJ, 313, 576
\bibitem[G\"urkan et al.(2004)G\"urkan, Freitag \& Rasio]{gurkan04}G\"urkan M.A., Freitag M., Rasio F.A., 2004, ApJ, 604, 632
\bibitem[Hamann \& Koesterke(1998)]{ham98}Hamann W.-R., Koesterke L., 1998, A\&A, 335, 1003
\bibitem[Heger et al.(2003)]{heger03}Heger A., Fryer C.L., Woosley S.E., Langer N., Hartmann, D.H.,	2003, ApJ, 591, 288
\bibitem[Heggie \& Hut(2003)]{gravmill}Heggie D.C, Hut P., 2003, The Gravitational Million-Body Problem, Cambridge University Press
\bibitem[Heggie et al.(1994)]{heggie94}Heggie D.C., Inagaki S., McMillan S.L.W., 1994, MNRAS, 271, 706
\bibitem[Heggie et al.(2006)]{heggie06}Heggie D.C., Trenti M., Hut P., 2006, MNRAS, 368, 677
\bibitem[Heggie(1975)]{heggie75}Heggie D.C., 1975, MNRAS, 173, 729
\bibitem[Hills(1980)]{hills80}Hills J.G., 1980, ApJ, 235, 986
\bibitem[Humphreys \& Davidson(1994)]{hump94}Humphreys R.M., Davidson K., 1994, Astronomical Society of the Pacific, Publications, 106, 1025
\bibitem[Hurley, Pols \& Tout(2000)]{hurley00}Hurley J.R., Pols O.R., Tout C.A., 2000, MNRAS, 315, 543
\bibitem[Hurley, Tout \& Pols(2002)]{hurley02}Hurley J.R., Tout C.A., and Pols O.R., 2002, MNRAS, 329, 897
\bibitem[Inagaki(1986)]{inagaki86}Inagaki S., 1986, PASJ, 38, 853
\bibitem[King(1966)]{king66}King I.R., 1966, AJ, 71, 64
\bibitem[Kroupa(2001)]{kroupa01}Kroupa P., 2001, MNRAS, 322, 231
\bibitem[Kudritzki(2002)]{kudritzki02}Kudritzki R.P., 2002, ApJ, 577, 389
\bibitem[Lamers et al.(2010)]{lamers10}Lamers H.J.G.L.M., Baumgardt H., Gieles M., 2010, MNRAS, 409, 305
\bibitem[Leitherer et al.(1992)]{leitherer&92}Leitherer C., Robert C., Drissen L., 1992, ApJ, 401, 596
\bibitem[Loup et al.(1993)]{loup93}Loup C., Forveille T., Omont A., Paul J. F., 1993, Astronomy and Astrophysics Supplement Series, 99, 291
\bibitem[Maeder(1992)]{maeder92}Maeder A., 1992, A\&A, 264, 105
\bibitem[Makino \& Sugimoto(1987)]{makino87}Makino J., Sugimoto D., 1987, PASJ, 39, 589
\bibitem[Makino(1996)]{makino96}Makino J., 1996, ApJ, 471, 796
\bibitem[Mapelli \& Bressan(2013)]{m&b13}Mapelli M., Bressan A., 2013, MNRAS, 430, 3120
\bibitem[Mapelli et al.(2013)]{map&13}Mapelli M., Zampieri L., Ripamonti E., Bressan A., 2013, MNRAS, 429, 2298
\bibitem[Martins et al.(2008)]{martins08}Martins F., Hillier D. J., Paumard T., Eisenhauer F., Ott T., Genzel R., 2008, A\&{}A, 478, 219
\bibitem[McMillan \& Engle(1996)]{mcmill96}McMillan S.L.W., Engle K.A., 1996, in Hut P., Makino J., eds, Proc. IAU Symp. 174, Dynamical Evolution of Star Clusters: confrontation of theory and observations, 379
\bibitem[McMillan(1986)]{mcmill86}McMillan S.L.W, 1986, ApJ, 307, 126
\bibitem[Muijres et al.(2012)]{muijres12}Muijres L., Vink J. S., de Koter A., Hirschi R., Langer N., Yoon S.-C., 2012, A\&{}A, 546, 42
\bibitem[Nanni et al.(2013)]{nanni13}Nanni A., Bressan A., Marigo P., Girardi L. 2013, MNRAS, 434, 2390
\bibitem[Pols et al.(1998)]{pols&98}Pols O.R, Schr\"oder K., Hurley J.R., Tout C.A., Eggleton P.P., 1998, MNRAS, 298, 525
\bibitem[Portegies Zwart \& McMillan(2002)]{port02}Portegies Zwart S.F., McMillan S.L.W., 2002, ApJ, 576, 899
\bibitem[Portegies Zwart \& Verbunt(1996)]{port96}Portegies Zwart S.F., Verbunt F., 1996, A\&A, 309, 179
\bibitem[Portegies Zwart et al.(2001)]{port01}Portegies Zwart S.F., McMillan S.L.W., Hut P., Makino J., 2001, MNRAS, 321, 199
\bibitem[Portegies Zwart et al.(2007)]{port07}Portegies Zwart S.F., McMillan S.L.W., 2007, in St-Louis N., Moffat A.F.J., eds., ASP Conf. Ser. Vol. 367, Massive Stars in Interactive Binaries, San Francisco, Astron. Soc. Pac., 597
\bibitem[Portegies Zwart et al.(2010)Portegies Zwart, McMillan, \& Gieles]{ymsc10}Portegies Zwart S.F., McMillan S.L.W., Gieles M., 2010, ARA\&A, 48, 431
\bibitem[Portinari et al.(1998)]{portinari&98}Portinari L., Chiosi C., Bressan A., 1998, A\&A, 334, 505
\bibitem[Sch\"oier et al.(2013)]{schoier13}Sch\"oier F. L., Ramstedt S., Olofsson H., Lindqvist M., Bieging J. H., Marvel K. B., 2013, A\&{}A, 550, A78
\bibitem[Schulman et al.(2012)]{schul12}Schulman R.D., Glebbeek V., Sills A., 2012, MNRAS, 420, 651
\bibitem[Sippel et al.(2012)]{sippel12}Sippel A.C., Hurley J.R., Madrid J.P., Harris W.E., 2012, MNRAS, 427, 167
\bibitem[Spitzer(1987)]{spitzer87}Spitzer L., 1987, Dynamical Evolution of Globular Clusters, Princeton University Press
\bibitem[Vesperini et al.(2009)]{vesp&09}Vesperini E., McMillan S.L.W., Portegies Zwart S., 2009, ApJ, 698, 615
\bibitem[Vink \& de Koter(2005)]{vink05}Vink J.S., de Koter A., 2005, A\&A, 442, 587
\bibitem[Vink et al.(2001)]{vink01}Vink J.S., de Koter A., Lamers H.J.G.L.M., 2001, A\&A, 369, 574

\end{thebibliography}
\end{document}